\documentclass[]{spie}  
\usepackage[]{graphicx}
\usepackage{amsmath}
\usepackage{url}

\title{Scanning strategies for imaging arrays} 

\author{Attila Kov\'{a}cs
\skiplinehalf
MPIfR, Auf dem H\"{u}gel 69, 53121 Bonn, Germany
}

\authorinfo{Further author information: E-mail: akovacs@mpifr-bonn.mpg.de, Telephone: +49 228 525 196}

\begin{document} 
  \maketitle 

\begin{abstract}
Large-format (sub)millimeter wavelength imaging arrays are best operated in scanning observing modes rather than traditional position-switched (chopped) modes. The choice of observing mode is critical for isolating source signals from various types of noise interference, especially for ground-based instrumentation operating under a bright atmosphere. Ideal observing strategies can combat $1/f$ noise, resist instrumental defects, sensitively recover emission on large scales, and provide an even field coverage -- all under feasible requirements of telescope movement. This work aims to guide the design of observing patterns that maximize scientific returns. It also compares some of the popular choices of observing modes for (sub)millimeter imaging, such as random, Lissajous, billiard, spiral, On-The-Fly (OTF), DREAM, chopped and stare patterns. Many of the conclusions are also applicable other imaging applications and imaging in one dimension (e.g.~spectroscopic observations). 
\end{abstract}



\section{INTRODUCTION}
\label{sec:intro}  

The latest generation of imaging arrays for submillimeter and far-infrared applications, both on the ground\cite{sharc2, gismo, laboca, scuba2, artemis} and in space\cite{spire, pacs}, produce total-power readouts, providing snapshot views not unlike optical and infrared cameras. This is in contrast to the instruments of the past where data consisted of difference signals from source and a nearby ``off-position''. It is true, that the differential readout scheme provides an effective way of rejecting atmospheric variations (for ground-based instruments), and/or other sources of $1/f$ type noise interference, provided that the differencing happens at a fast enough rate. In practice, however, the noise rejection is rarely perfect (resulting in ``striping'') and the reconstruction of images\cite{emerson88, emerson95, wright} from differenced data is at once challenging and riddled with problems due to the inadvertent filtering of spatial scales. Nevertheless, differencing remains the only effective way of observing from the ground, with a bright and variable atmospheric foreground, with just a single or a few pixels.

However, the large imaging arrays of today can do better since they collect information at many positions simultaneously, hence no longer needing explicit differencing of ``on'' and ``off'' positions. The noise (e.g.~correlated atmospheric variations, and detector $1/f$ noise) can be effectively separated from the astronomical source signals by capable algorithms\cite{crush}, provided that the source signals are ``moved'' from pixel-to-pixel during the observation. However, sensitive recovery is only possible for source components, which do not overlap with the predominant noise signals. Some modes of observing are inherently better in isolating the source signals from noise interference and providing sensitivities to more extended spatial scales.

The benefits of faster scanning and that of cross-linking have been widely recognised\cite{fastscanning, weferling, tegmark, waskett}. Faster scanning moves the source signals into the higher frequencies of the detector signal spectra, where the $1/f$ interference is less. Cross-linking assures that all Fourier components of the source, along all spatial directions, are scanned at the higher speeds. Several patterns are now used or have been proposed for various instruments. However, some of the observing modes are objectively better than others. Earlier attempts at systematic comparisons explored only some of the aspects involved\cite{tegmark}, or described a framework but did not actually compare observing strategies\cite{thesis}. This paper aims to provide the most comprehensive evaluation yet of observing strategies for (sub)millimeter-wave imaging.

In the first part of the paper, we outline the criteria that make good observing modes and provide guidelines for designing these. In the second part, we use this understanding to define quantitative measures that are consequently used on a selection of commonly used, or suggested, imaging observing patterns.

\section{DISCUSSION}
\label{sec:discussion}

Before jumping into analyzing different scanning modes, one must first ask what are the qualities we actually seek in an observing mode. Here are some generic considerations for astronomical applications:

\begin{itemize}

\item[] {\bf Resistance to Noise.}
Perhaps the most critical aspect of a good observing mode is its ability to stand up to various sources of adverse noise interference. Some of the noise may be expected (e.g.~$1/f$ type noise from electronics or sky-noise) while others may be from more unforeseen sources (e.g.~electronic pickups in the telescope environment). Scan patterns that are versatile in their ability to take on noise of arbitrary types are more robust, and therefore more desirable.

\item[] {\bf Large-Scale Sensitivity.}
In astronomy, the nature of the underlying emission is often extended and heavily resolved by the telescope beams. The accurate and sensitive recovery of the large scales is often pivotal to the scientific goals of the observation. Thus, scans patterns that facilitate this are at an advantage.

\item[] {\bf Coverage.} 
Uniform coverage (i.e.~sensitivity) over the full area of an observed field is generally desired, and scan patterns should provide it as much as possible.

\item[] {\bf Signal Dynamic Range.}
Provided that sensitive measurement of faint signals is the very goal of observation, one should in general avoid observing modes that result in large jumps in signal power (e.g.~due to changes in the optical background, which could drown out the low-lying information of interest.

\item[] {\bf Feasibility.} 
Last, but not least, one ought to consider the physical limitations on what scanning modes could be implemented. Modes that move entire telescopes will be limited to tracing smooth patterns within the acceleration limits of telescope drives. Patterns using secondary wobblers are similarly limited in the settling-time required to move between positions, as well as in their range of movement.

\end{itemize}

These are the most important typical criteria that should enter into the evaluation of specific scanning patterns. Other criteria may be applied to specific experiments (e.g.~ability to accurately measure a spatial spectrum for certain CMB experiments\cite{tegmark}). At the same time some of the criteria listed here may become unimportant for other purposes (e.g.~the recovery of large-scales may be irrelevant for deep-field surveys focusing on compact sources). 

\subsection{Noise Isolation}
\label{sec:noise}

\begin{figure}[!htbp]
\centering
\includegraphics[width=0.4\textwidth]{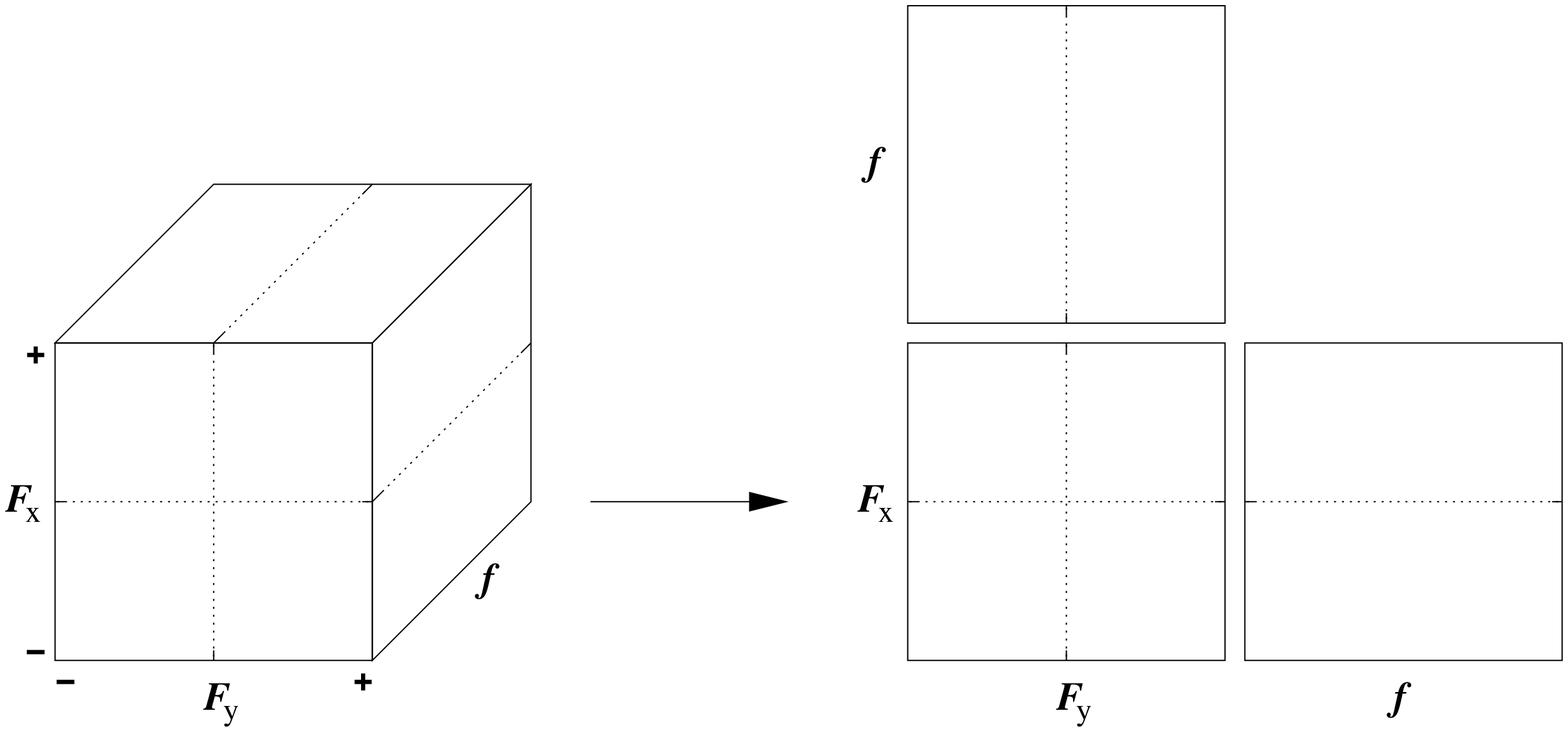}
\caption{The projections of a spectral cube.}
\label{fig:unfolding}
\end{figure}

Resistance to the harmful effects on noise is one aspect which observing modes can almost single-handedly control. The most common type of adverse noise is stationary, both along physical directions and in time. Such noise, is best described by its power spectrum of independent components, obtained by Fourier transforming signals along the imaging coordinates and time. Thus, the commonly distracting $1/f$ type noise resides mostly in the low frequencies, while narrow and wide-band resonances occupy distinct regions in phase-space. With two-dimensional imaging arrays sampled in time, both the data cube, and its power spectrum are three-dimensional. Its symmetry allows us to consider only the positive temporal frequencies, but retaining both positive and negative spatial frequencies at the same time. We can most easily visualize such spectra by the plane-projections along the principal directions (see the ``unfolding'' of a spectral cube in Fig.~\ref{fig:unfolding}).

\begin{figure}[!htbp]
\centering
\includegraphics[width=0.3\textwidth]{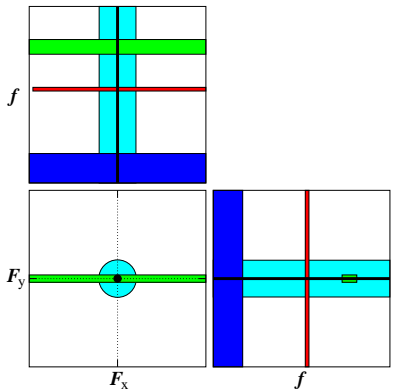}
\caption{Example illustration of the typical spectral locations of noise. Shown are correlated noise (thick black line), as well as higher-order sky-noise, detector $1/f$ noise, and resonances (both an isotropic narrow-line, and oriented wide-band resonances in some or all detector columns) shown as shaded areas. }
\label{fig:noise-locations}
\end{figure}

Such a view is useful in identifying the common spectral locations of noise (Fig.~\ref{fig:noise-locations}) that can affect astronomical observing. The canonical $1/f$ noise from detector drifts affects the low temporal frequencies for all spatial components, with the lowest frequencies being noisiest. Correlated noise (such as $1/f^2$ temperature drifts, or background opacity variations) will typically pollute all temporal frequencies along the zero spatial frequency component. Higher-order sky noise can affect some low spatial-frequency components around that. Resonances are restricted to specific temporal frequencies, but can vary in the spatial distributions. Some of the noise may come as common mode signals, which can spread along all spatial directions, while others will have a particular orientation (e.g.~correlated detector columns affecting all frequency components in the $x$ frequency direction, but not in $y$, (e.g.~the electronic drifting of SHARC-2 detector rows).

\subsubsection{Orientable Noise Patterns}

Such orientable noise patterns are particularly interesting from the perspective of observing-mode design. If a mode of scanning can rotate the source spectrum with respect to another strategy, then some of the source components falling into the region of directional noise in one scan will be clear of interference in the other and vice versa. Thus, the combination of two or more such scans can effectively bypass the harmfulness of directional noise. The realignment can happen either by explicit instrument rotation (where this is possible), or in case of ground-based telescopes, by the field-rotation as the source transits across the sky. Coherent analysis\cite{crush, thesis} of the rotated patterns can later fully recover the source signals from directional noise interference.

\subsection{Large-Scale Structures}
\label{sec:large-scales}

Large-scale structures tend to be more challenging to observe. For data acquisition modes that move along connected trajectories (vs.~discrete pointings), it takes longer to move detectors across extended structures than compact ones, mapping the source flux into the lower frequencies of the signal spectrum, where $1/f$ type noise interference is more severe. For imaging arrays, this means that the extended source signals tend to concentrate in just a small volume ($\sim$$1/L^3$) of the available phase space, making these much more exposed to the harmful effects of noise interference in general. To obtain the effective signal spectra of extended Gaussian-shaped sources scanned across with a typical speed $v$, one can simply multiply the ideal point-source spectra (see Sec.~\ref{sec:simulations}) by a Gaussian taper with respective standard deviations of $\sigma_F \approx 0.37/{\rm FWHM}$ in the spatial directions and $\sigma_f \sim 0.37 v/{\rm FWHM}$ in the time direction. Thus, it is sufficient to provide spectra of perfect point sources for a given scanning pattern for characterizing its phase-space behaviour for source structures of arbitrary scales.

\begin{figure}[!htbp]
\centering
\includegraphics[width=0.4\textwidth]{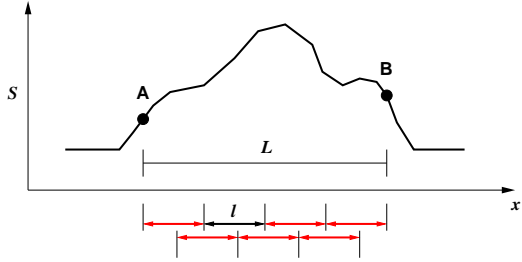}
\caption{Recovery of the extended emission.}
\label{fig:extended}
\end{figure}

The accuracy with which large scales can be measured is another important issue. Consider Figure~\ref{fig:extended}: the recovery of the extended emission on the scales $L$ translates to recovering the difference in fluxes $S$ between two points $A$ and $B$ separated by $L$. When both points are observed with the same detector during its $1/f$ stability time-scale, the comparison is straightforward and limited only by the detector's white noise level. However, when the observing mode moves detectors in a smaller range $l$ during the critical time-scales of the detector stability, the large-scale flux comparison becomes more tedious. Several ``intermediate'' flux comparisons have to be made, using overlapping regions spanned by detectors, before the measurement of point $A$ by one detector can be related to the measurement of $B$ by another detector. Each of the intermediate comparisons add extra contributions to the total uncertainty of the desired difference measure. It is easy to convince oneself that the effective noise rms $\sigma_L$ on scales of $L>l$, grows as

\begin{equation}
\sigma_L \approx \left( 1 + \frac{L}{l} \right)^{1/2} \sigma_0
\label{eq:extended-noise}
\end{equation}

in terms of the scanning range $l$ (spanned within the $1/f$ stability time-scale), and the compact source (i.e.~white-noise) limit $\sigma_0$.

Another practical limit on the large scales is imposed if the detector array operates under strong correlated signals (e.g.~atmospheric variations, or correlated electronic noise on groups of pixels). Often, no meaningful sensitivities can be reached unless such correlated signals are first removed. However, with the correlated noise go also the scales larger than those affected by the correlations. For ground-based submillimeter detectors, which operate under a bright and variable atmosphere, this normally means that structural scales larger than the array field-of-view (FoV) are usually lost\cite{crush}. Often, the limiting scale to sensitive measurements is yet smaller due to correlated groups of detectors within the array.

\subsection{Design Considerations}
\label{sec:design}

Based on the discussion thus far, we can already outline some guidelines as to how to design better observing strategies. 

Harmful noise interference may come in any spectral component. Therefore, observing modes which spread the source signals more widely in phase space are more prone to be immune in general. There is also advantage to occupying more of the higher-frequency components both in spatial and temporal directions as these are typically less influenced by $1/f$ type noise. Somewhat unfortunately, any mode of observing whose time stream source signals have definite signs (positive or negative), will always have most of the source signal power at the lower frequencies. Accordingly, the high-frequency regions of phase space cannot be selectively occupied, but flux can at least be spread into these far enough such that the low-frequency information loss due to noise is less critical. The faster the scanning, covering more positions in a given time, the higher its span in temporal frequencies.

The most uniform spreading of source signals is achieved by stochastic observing modes (these provide the only viable way of spreading source simultaneously in phase-space and in time where the source must be observed many times over). Since truly random patterns are often not feasible for the fast positioning of telescopes, strategies with a random-like appearance (with quasi-random source crossings in the detector time-streams), are expected to perform best.

To provide optimal sensitivities on the larger scales, it is necessary that the observing mode spans, within the $1/f$ detector stability time-scales, distances that compare in size to the extended structures to be measured. In case of 2-D imaging, this requires truly 2-D patterns with sufficient reach in both directions. 2-D scanning is essential also for spreading the source signals into the available volume of the phase space.

Signal dynamic range considerations have implications for observing under strong backgrounds (e.g.~ground-based submillimeter telescopes). In general, scanning can be implemented either by moving entire telescopes (i.e.~the primary reflector), or with the smaller secondary mirror (or another mirror further down the optical path). While small secondaries are pushed around more easily and can be positioned faster than entire telescope structures, the agility comes at a steep price. Moving mirrors down the optical path alter the illumination of the primary reflector by the instrument. Whereas this may be no major concern in low-background environments (e.g.~space or air-borne telescopes, optical applications), it becomes an issue in ground-based submillimeter observing where backgrounds tend to trump faint astronomical signals by several orders of magnitude. The residual background signals produced by the illumination changes from secondary motion can be large, and difficult to model accurately enough in the analysis. Note, that while this problem is overcome (albeit imperfectly resulting in ``striping'') by symmetric ``chopping'' when differential readouts are used, it is not nearly as easily handled in total-power modes. Therefore, scanning modes using primary reflector motion are generally cleaner and preferred for ground-based submillimeter applications.

In summary, these are the guiding principles for sound observing strategies:

\begin{itemize}

\item{Faster is better}.

\item{2-D scanning for 2-D imaging.}

\item{Random or quasi-random source crossings.}

\item{Spanned range should match the largest observed scales.}

\item{Scanning with primary reflector preferred (for ground-based submillimeter telescopes).}

\end{itemize}

These are quite general principles, which can be applied (with some adjustments) to observing strategies outside the realm of astronomical imaging also.

\section{ANALYSIS}
\label{sec:simulations}

We simulated{\footnote{The {\em JAVA} source code of the simulations is available at \url{http://www.submm.caltech.edu/~sharc/scanning}. It can be used to evaluate (arbitrary) imaging observing modes, which are not discussed in this paper. Readers should feel free to download and use this software to test performance of their observing modes of choice.} various typical imaging observing modes on a 32$\times$32 pixel (i.e.~$\sim$1 kilo-pixel) filled test-array. The scanning patterns (see Sec.~\ref{sec:patterns} further below) were used to place a mathematically perfect point source, at the tracking center, into one of the array pixels for each frame. The resulting data cube was then Fourier-transformed ($F_{t, {\bf x}} \rightarrow \tilde{F}_{f,{\bf F}}$) along all three dimensions, in time windows of 64 frames, to obtain the averaged phase-space power spectrum ($P \sim |\tilde{F}|^2$) of a given scanning pattern. The power spectra were subsequently renormalized ($\hat{P}$) to their peak-value at non-zero spatial and temporal frequencies. The omission of the zero frequencies in the normalization reflects that fact that the zero-bin components are typically drowned in $1/f$ noise (see Sec.~\ref{sec:noise}). The plane-projections (cf.~Fig.~\ref{fig:unfolding}) of the thus normalized power spectra are shown on Fig.~\ref{fig:spectra}. To measure the noise resistance of scanning patterns, we define their phase space moments:

\begin{equation}
m_i = \left< f^i \hat{P}_{f, {\bf F}} \right> = \frac { \sum_f \sum_{\bf F} f^i \hat{P}_{f, {\bf F}} } {  \sum_f \sum_{\bf F} f^i}.
\end{equation}

The zero-order moment ($m_0$) provides a measure of the volume fraction occupied in phase-space by the pattern. As such, it is an indicator how immune a pattern is to noise interference that may come from any region in frequency-space. The moments $m_1$ and $m_2$ downweight the lower temporal frequencies exactly as would be appropriate for pure $1/f$ or $1/f^2$ noise respectively. Thus these higher moments serve as measures of the resistance provided by the observing pattern against $1/f$ and $1/f^2$ type noise interference. The higher moments are also qualitative indicators of typical noise resistance at the more extended source scales, whose spectra are restrained by a taper, which suppresses the higher frequencies for scanning modes (see Sec.\ref{sec:large-scales}).

To avoid comparing apples to oranges, the various scan patterns were uniformized as much as possible. Mean scanning speeds (where applicable) were set at 1 pixel/frame, allowing patterns to reach all of phase-space up to the Nyquist frequencies of the simulation. By the same argument, discrete patterns (such as {\em random}, {\em chop} or {\em DREAM}) were simulated with 1 position/frame ``movement''. The size of patterns were also fixed (as much as possible) to cover the same areas. The simulations were mostly performed for mapping small, FoV-sized fields. As a compromise between uniformity of coverage and sensitivity to large scales, scan patterns were aimed to fit inside a 16$\times$16 pixel area (i.e. a quarter of the test array). This was adjusted to slightly oblique shapes for {\em billiard scans}, while {\em chop} modes simply used 16-pixel throws in the chopping direction(s). The size of the {\em DREAM} pattern is fixed by design. The resulting comparison among patterns is therefore reasonably fair. The results are summarized in Table~\ref{tab:summary} at the end of this paper.

Most of the scan patterns (except {\em random}) occupy fractal surfaces in phase-space representations. This means that the absolute values of the moments depend on the choice of spectral resolution for the simulation (the more frequency bins used, the lower the calculated moments will become). However, our simulations showed that the spectral resolution does not affect the relative results between non-random patterns in a significant way. Thus, for the simulations we used the minimal resolution of 32 frequency channels (hence the 64 frame time windows) required to maintain the full dynamic range of source frequencies, from point source to FoV, as they are scanned by the patterns with 1\,pixel/frame (or 1 position/frame) mean scanning speeds. For the pictorial representation of spectra (Fig.~\ref{fig:spectra}) we used a higher resolution to provide a better glimpse at the structural details involved.

\section{OBSERVING MODES}
\label{sec:patterns}

%

We examined some of the commonly used, or proposed, observing modes for submillimeter imaging with large arrays. By no means is this list exhaustive, but it is nevertheless a representative sample of the types of observing strategies commonly used or proposed at present. We included both modes consisting discrete pointings (random, chopped and DREAM) and smooth, patterns with continuous movement, some of which provide acceleration-free travel times (OTF and billiard), while others offer moderate curvatures (Lissajous and spirals). Some are oriented patterns while others are truly two-dimensional. Nevertheless, should the list of patterns be lacking, the same comparative evaluations are easily extended to other observing modes of choice.

\subsection{Random Positions}

As already mentioned, random observing patterns are expected to be best, since they spread source signals, present often or at all times in the array field-of-view, most uniformly into phase-space, thus providing the best possible immunization against the harmful effects of noise. However, random positioning is not easily realized fast enough with mechanical movement. It may beyond hope for most telescopes and submillimeter applications where stability time-scales may be short\footnote{Although, when stability allows, random modes can be possible. An example is provided by the APEX\cite{apex} {\em hexa} observing mode for the CHAMP+ spectroscopic receiver array. The small span of the positions and the relatively long time exposures (of seconds) per pointing allow a random pattern of telescope pointings to be realized.}, except perhaps by using moving secondaries (or optics further down in the optical path). However, the moving of optical elements other than the primary reflector can be harmful otherwise (see Sec.~\ref{sec:design}). Nevertheless, random patterns may be well-suited to some other applications, such as scanning in frequencies for innovative spectroscopic observing modes.

\begin{figure}[!htbp]
\centering
\includegraphics[width=0.3\textwidth]{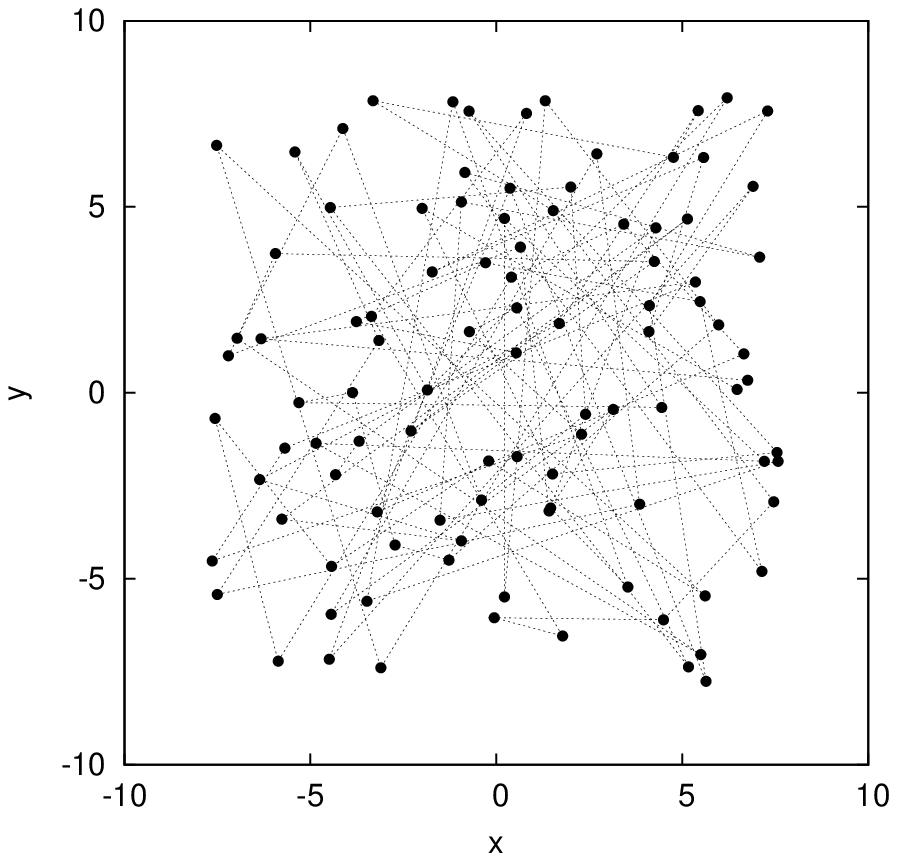}
\includegraphics[width=0.3\textwidth]{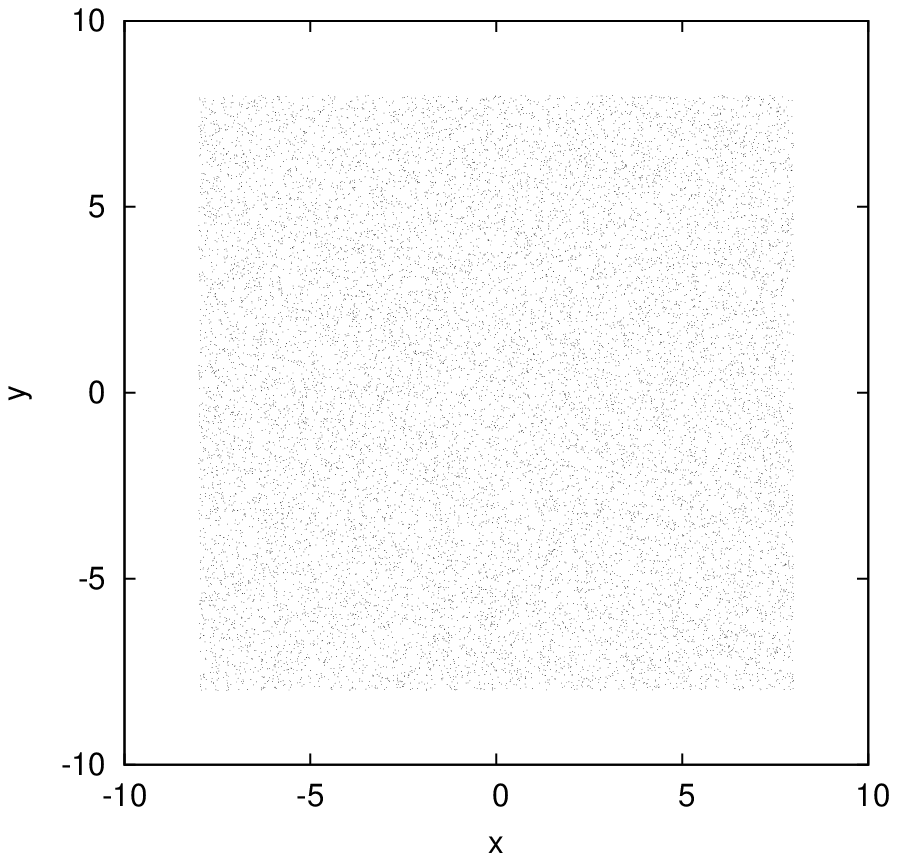}
\caption{Random Positions. A snipplet of the pattern (left), and the positions covered after some time (right).}
\label{fig:random}
\end{figure}

\subsection{Stare Modes}

Despite all the arguments presented already on the need of moving the source signals among pixels, it is necessary to digress a little and briefly mention what happens if one does not abide by that but decided instead to ``stare'' at a single position for some time, like optical cameras often do. Such mode of observing is one of the main modes envisaged for operating SCUBA-2\cite{scuba2}. While it is recognized that the mode is very much limited by the $1/f$ drifts of detectors, it is meant to be countered by the occasional taking of calibrating ``dark frames''. However, even so the mode has serious handicaps when compared to any of the scanning modes.

Even if the system were entirely stable under a full ``sky--dark'' cycle (and that is already a generous assumption), the error $\sigma_S$ of the resulting source estimate from $N$ frames of total observing is given by

\begin{equation}
\sigma_S = \frac { \sigma_{\rm det} } {\sqrt{N}} \left ( \frac{1 + r^2 } { 1-\alpha } + \frac{ 1 } { \alpha } \right)^{1/2}
\label{eq:stare}
\end{equation}

for the instrumental (i.e.~dark-frame) detector noise $\sigma_{\rm det}$ in a single frame, a relative on-source background noise (i.e.~photon shot-noise) $r = \sigma_{\rm bg} / \sigma_{\rm det}$, and dark integration fraction of $\alpha$. Accordingly, minimum measurement error is reached when 

\begin{equation}
\alpha = \frac { \sqrt{1 + r^2} - 1 } { r^2 }. 
\end{equation}

Clearly, the dark-frame integrations become negligible only in the limit $\sigma_{\rm bg} \gg \sigma_{\rm det}$. This may well be true for the typical optical or infrared detectors. However, the ground-based submillimeter arrays are only marginally background limited with $r \approx 1$, hence requiring dark frame integrations that approach on-source observing times. For detector-noise limited airborne or space-based instruments, or mm-wave detectors $r \approx 0$, and therefore stare mode observing in low backgrounds require dark-frame observing half of the time. Furthermore, dark frames not only require time to acquire, but the differencing with them also increases the noise of the source estimate (Eq.~\ref{eq:stare}) when compared to the noise reached in pure on-source times by scanning modes. These combined overheads can become significant (up to 4-times in integration time to reach the same rms as scan patterns), and thus render stare modes impractical for submillimeter applications.

An even more compelling argument against staring is provided by the drastic change in optical loading between on-sky and dark frames, which produce signals often $\sim$$10^6$ times brighter than those from faint astronomical sources in ground-based submillimeter applications. Under such conditions, any small nonlinearities of the detector response larger than 1\,ppm could prove a show-stopper for deep integrations. Detectors are hardly ever that linear in truth. Thus, while stare modes find application in optical astronomy, where detectors are quiet and source signals are not too much fainter than the typical backgrounds, this mode is to be avoided in submillimeter astronomy whenever sensitive imaging is required.

\subsection{Position Switching (Chopping)}

The simplest true ``scanning'' mode is when the array is alternately pointed between two positions. The separation $d$ of these two positions can be smaller than the field-of-view, in which case an area in the center of the array remains observed all of the time, providing maximum sensitivities therein. However, as with all modes which consist of discrete positions, the fast switching that is required for $1/f$ type noise rejection (which may or may not include the very variable atmospheric foreground) may be difficult to achieve. For ground-based submillimeter applications, the fast switching, without too much overhead, can be realized using a moving secondary -- very much like in the case of the differential chopping observing mode. However, instead of differencing signals as before, the total-power data from those positions can be reduced the same way as for any other scanning mode\cite{crush}. 

The obvious problem with chopping as a total-power scanning mode is that if it is realized via secondary mirror movement, the illumination change of the primary will produce large signals that trump faint sources signals by orders of magnitude when done under a bright atmospheric foreground. This makes the accurate and sensitive recovery of source signals difficult to the point of practically impossible. Thus chopping is not ideal for ground based applications, though it may find use in low-background environments, such as airborne or space-based facilities.

Besides, chopping encodes the source signals at just one non-zero frequency (the chopping frequency), making very little use of the available phase space volume. As a result, its resistance to various types of noise is limited, though it provides excellent isolation from $1/f$ type interference. It is also an essentially 1-D pattern, thus strongly directional by nature. Unless cross-linked at an angle (preferably 90$^\circ$), or several angles, it is not very useful for the uniform recovery of large-scales along different spatial directions. Note, that cross-liked chopped scans produces the same moments as 1-D chops, which is why these are not listed separately in Table~\ref{tab:summary}.

\subsection{DREAM}

The Dutch Real-time Acquisition Mode\cite{dream} (DREAM) has been proposed as one of the principal observing patterns for the 10-kilopixel SCUBA-2 array\cite{scuba2}. It is a cyclical pattern consisting of a sequence of 16 discrete pointings on nearby pixels. The argument for this pattern is that by using a secondary wobbler it allows fast switching between positions while keeping most of the field-of-view under constant exposure.

\begin{figure}[!htbp]
\centering
\includegraphics[width=0.3\textwidth]{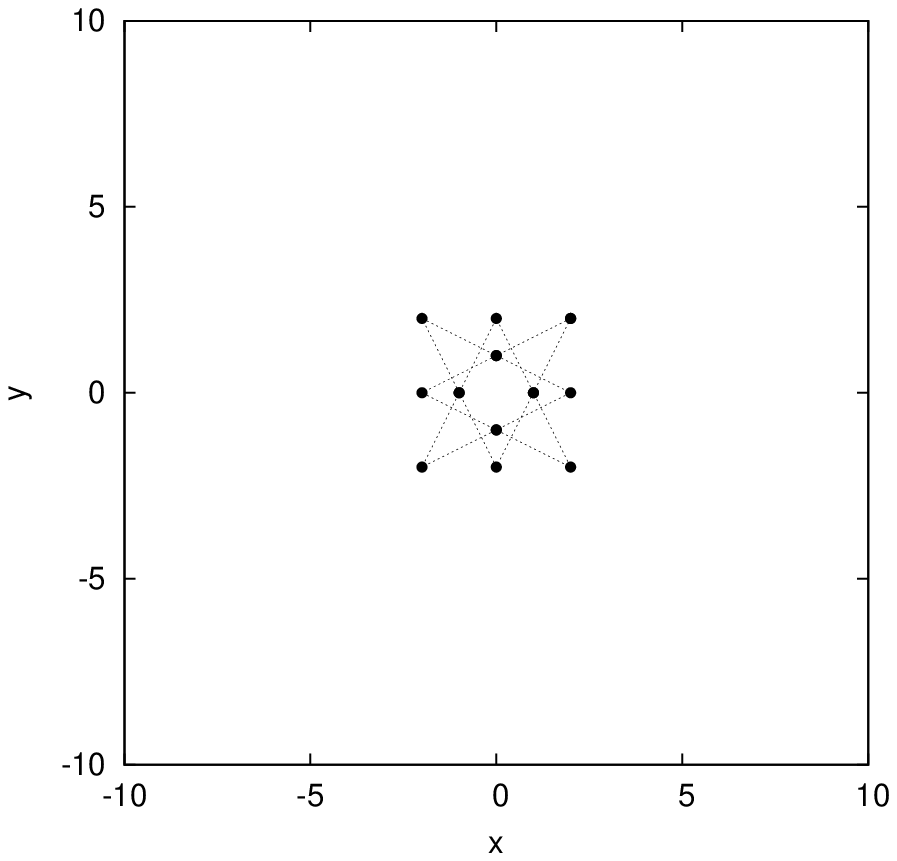}
\caption{DREAM}
\label{fig:dream}
\end{figure}

Yet, therein lie some of the obvious the problems of the pattern. As we already pointed out, the use of the secondary mirror for scanning is highly troublesome due to the changing illumination of the primary mirror with off-axis secondary positioning. This will produce strong DC offsets in the pixel time-streams between positions as residues of the bright atmospheric foreground. Accurate removal of these in the analysis is likely beyond hope for observing faint astronomical sources that are many orders of magnitude below the foreground signal powers. Besides, with is small span of just 4$\sqrt{2}$ pixels separating the farthest positions, the pattern is also rather limited in its ability to recover faint extended emission. Noise will quickly accumulate on the larger scales, with scales near the SCUBA-2 FoV being approximately 4-times noisier than the compact sources of $1-2$ beams across. Moreover, the periodic nature of the pattern means that it encodes source signals in just a few (9) discrete frequencies, providing very little isolation from possible noise interference. Some consolation is to be offered by the fact that these 9 frequencies spread across the available phase space -- thus providing some immunity against the most common $1/f$ type noise interference.

In summary, while its name would suggest otherwise, the DREAM mode is very far from ideal. It usefulness is probably limited to bright compact sources (e.g.~pointing), but these are also more effectively observed with other patterns.

\subsection{On-The-Fly (OTF) Mapping}

On-The-Fly mapping (also called ``serpentine''\cite{tegmark} pattern or ``raster scan''\cite{waskett}) has a long history in astronomy. The motion is back-and-forth along alternating rows. Each row has length $L$ and are spaced $\Delta$ apart. Thus, the motion is essentially linear at a constant speed, for most of the travel time, requiring accelerating only at the turnarounds. The pattern is attractive for its simplicity and because of the even field coverage it produces for arbitrarily large patterns.

\begin{figure}[!htbp]
\centering
\includegraphics[width=0.3\textwidth]{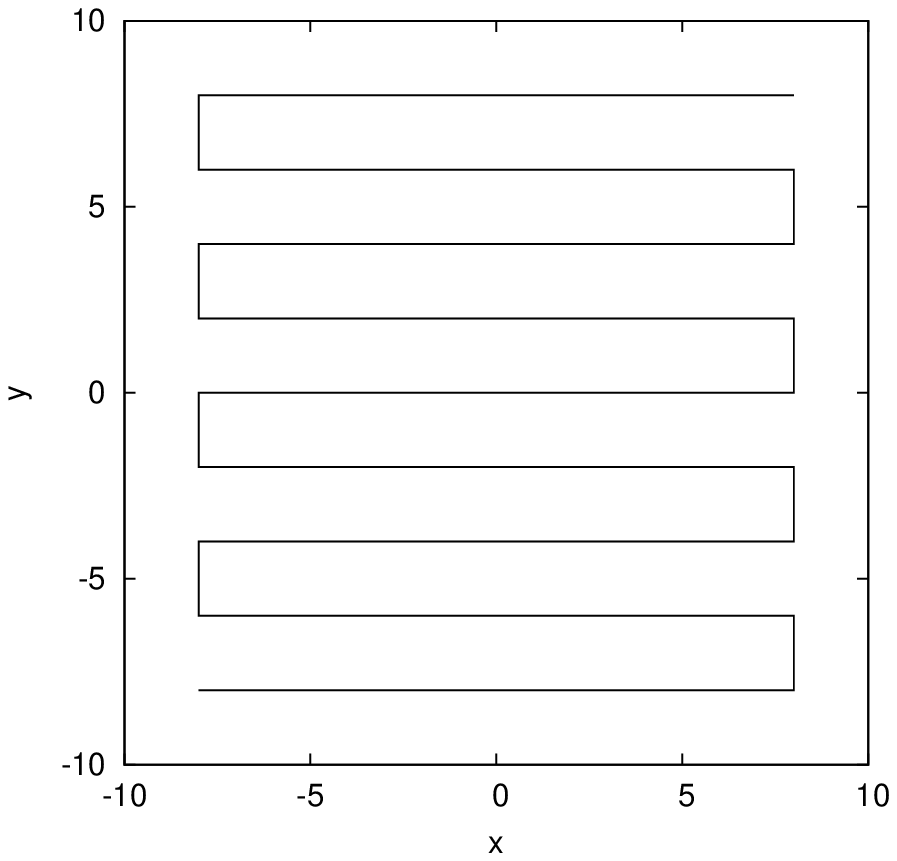}
\includegraphics[width=0.3\textwidth]{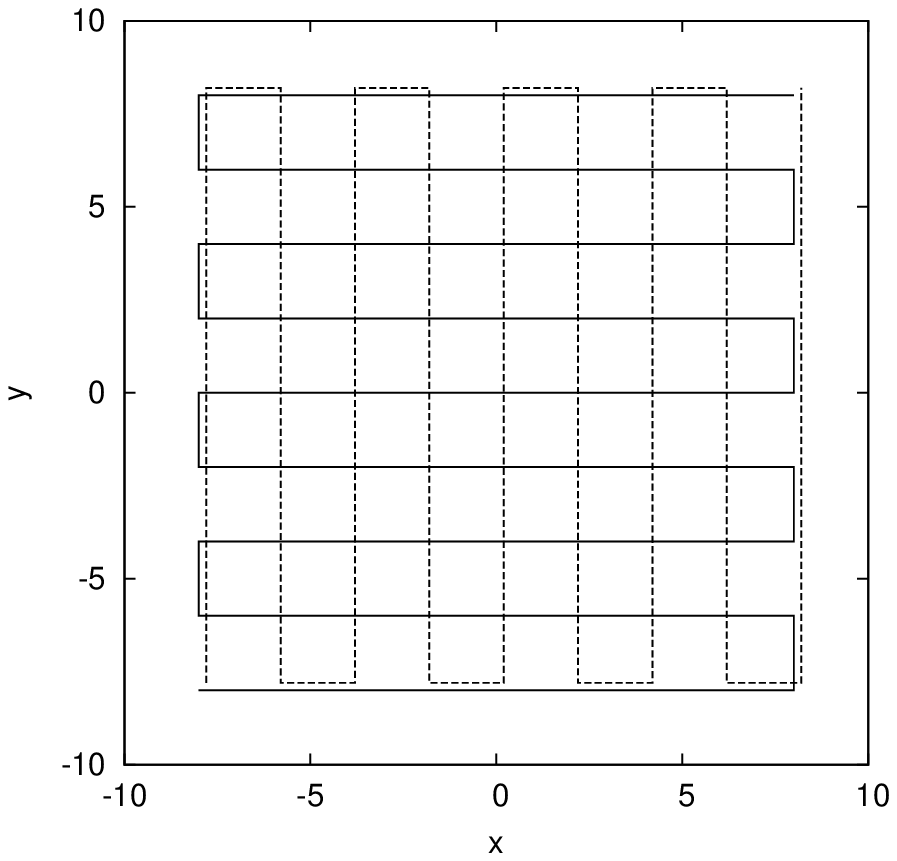}
\caption{On-The-Fly (OTF), single pass (left) and 90$^\circ$ cross-linked (right) with the orthogonal pattern dashed and slightly offset for visibility.}
\label{fig:otf}
\end{figure}

However, the fast motion happens along the direction of the rows only, while the motion between rows is at best a crawl ($v_y/v \sim \Delta / L$). Accordingly, one should expect strong directional sensitivities and noise resistance in the two orthogonal directions. This is confirmed by the simulations, which make such patterns very undesirable.


The strongly oriented nature of the OTF pattern is easily overcome by cross-linking two or more of these patterns at angles to one-another, such that the fast scanning along the rows can benefit all spatial directions. Optimal cross-linking strategies should span the full range of angles. For the simulations we used patterns that were cross-linked at the maximal $\delta \Theta = 90^\circ$ angles ({\em crossed OTF}), as well as a set spanning 0--180$^\circ$ in 10$^\circ$ steps ({\em rotated OTF}). As expected, the resulting source signals from crossed and rotated OTF scans better fill the available phase space in both directions. Even so, most of the source signal power remains constrained to relatively few spectral components when the crossing is at a single (90$^\circ$) angle. Cross-linking further addresses the otherwise very directional sensitivity to large scales of single OTF patterns, providing more-or-less uniform sensitivity to extended structures in all directions.

\subsection{Lissajous}

Lissajous patterns have been first developed for submillimeter observing in 2002 by the author for use with the 350\,$\mu$m SHARC-2 array. It is the main observing mode for that camera, and has since been adapted elsewhere also (e.g.~ASTE\cite{aste}). Lissajous patterns are governed by the independent oscillatory equations:

\begin{align}
x &= A_x \sin(\omega_x t + \phi_x), \\ 
y &= A_y \sin(\omega_y t + \phi_y).
\end{align}

The shape and coverage of the pattern is determined by the ratio of angular frequencies $\omega_y / \omega_x$ in the two directions. Frequency ratios that are not expressed as simple fractions give full coverage within the enclosing 2$A_x$$\times$2$A_y$ box size. The best ratios are irrational (resulting in non-repeating open patterns with perfect coverage over long time scales) and are of order unity in magnitude for providing similar effective scanning speeds in both the $x$ and $y$ directions. In the simulations we used a frequency ratio of $\sqrt{2}$, while keeping mean scanning speeds at 1 pixel/frame, in line with our stated guidelines.

\begin{figure}[!htbp]
\centering
\includegraphics[width=0.3\textwidth]{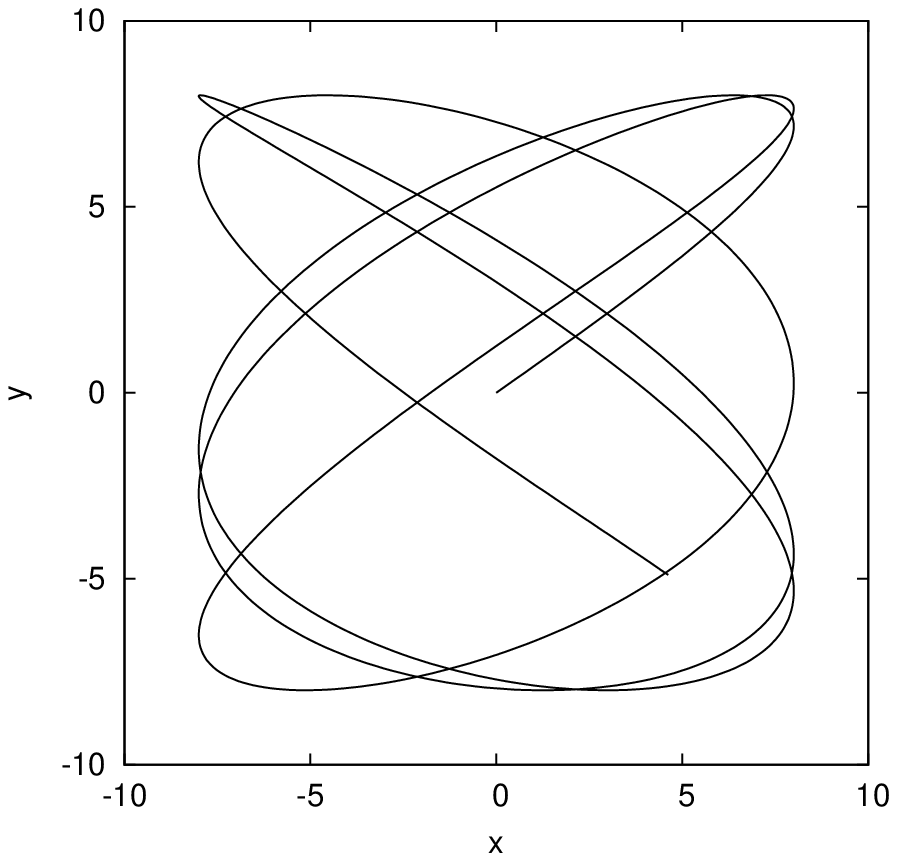}
\includegraphics[width=0.3\textwidth]{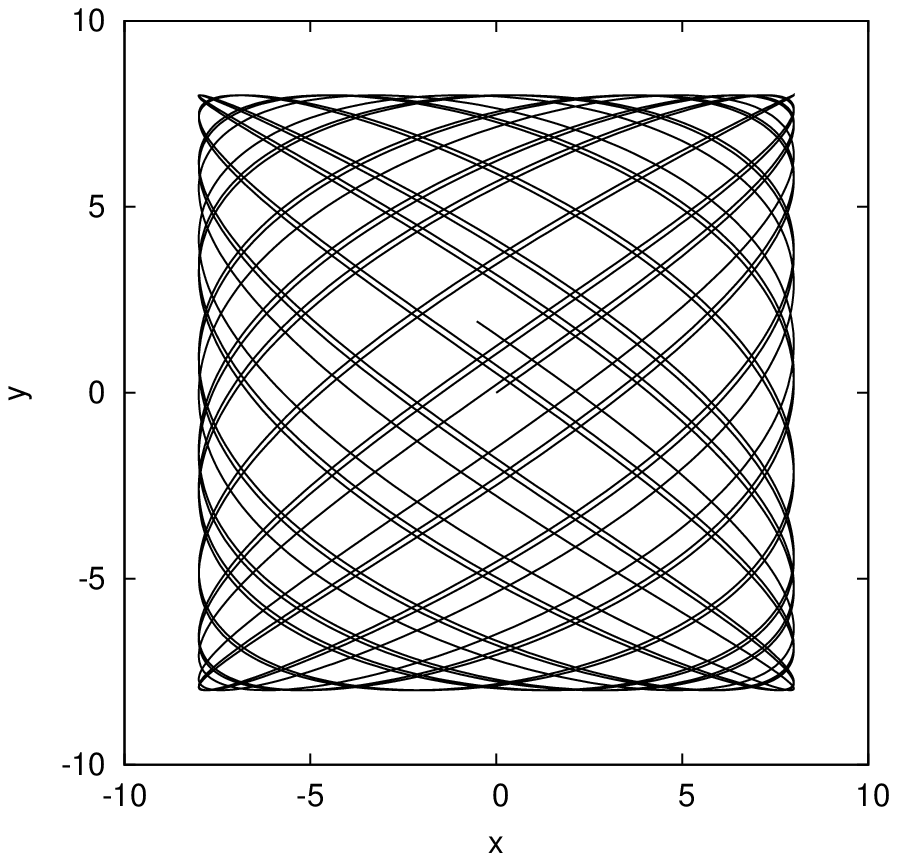}
\includegraphics[width=0.3\textwidth]{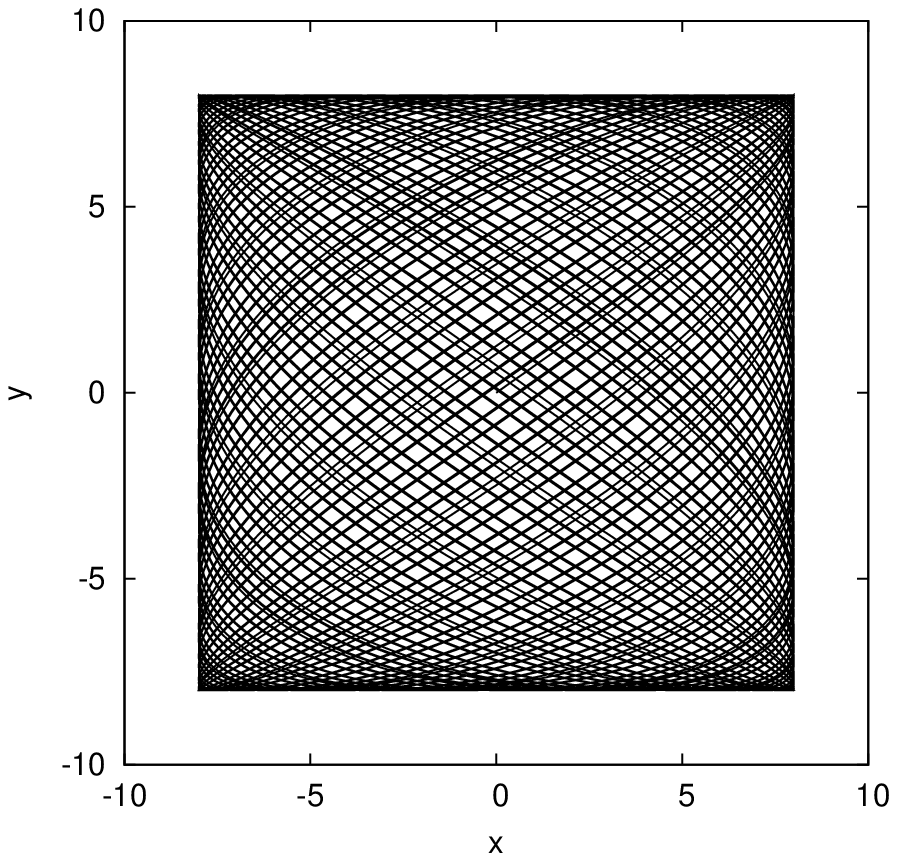}
\caption{Lissajous Pattern. At the beginning (left) and after longer integrations (center and right).}
\label{fig:lissajous}
\end{figure}

The pattern is perhaps the most attractive one for several reasons. The non-repeating nature effectively ensures quasi-random source crossings in the detector time-streams, thus mimicking random patterns to a certain degree. The result is a nearly uniform spreading of the source signals in phase space, which provides the best resistance against both generic and $1/f$ type noise interference among all the non-random patterns investigated. In addition, the pattern is smooth at all times without sudden changes in direction, thus requiring modest mechanical accelerations for implementation. It is also scalable in size, providing control over the trade-off between uniform FoV coverage and sensitivity to large scales.

Its only notable weakness is that Lissajous patterns ``spend'' too much time on the edges relative to the center of the pattern. For patterns larger than the field-of-view this may become bothersome. For smaller, FoV-sized mapping, however, this small defect turns into a slight advantage since the longer exposures over the edges partially compensate for the fewer pixels passing over those areas. Thus, Lissajous patterns are best-suited for observing small fields near FoV scales.

\subsection{Bouncing Billiard Ball}

This scanning pattern was originally developed\footnote{\url{http://www.submm.caltech.edu/~sharc/operating/boxscan.htm}} in 2002 by C.~Borys and C.D.~Dowell for covering large fields with SHARC-2. They named it ``box'' scan after the bounding rectangular box in which the pattern resides. The name ``pong'' scan has been used more recently in relation to SCUBA-2\cite{scuba2}. The pattern essentially emulates the bouncing of a ball on the billiard table (hence the name we use here), reflecting on edges whenever these are reached, but moving along straight trajectories and at an angle (usually 45$^\circ$) to the sides otherwise.

\begin{figure}[!htbp]
\centering
\includegraphics[width=0.3\textwidth]{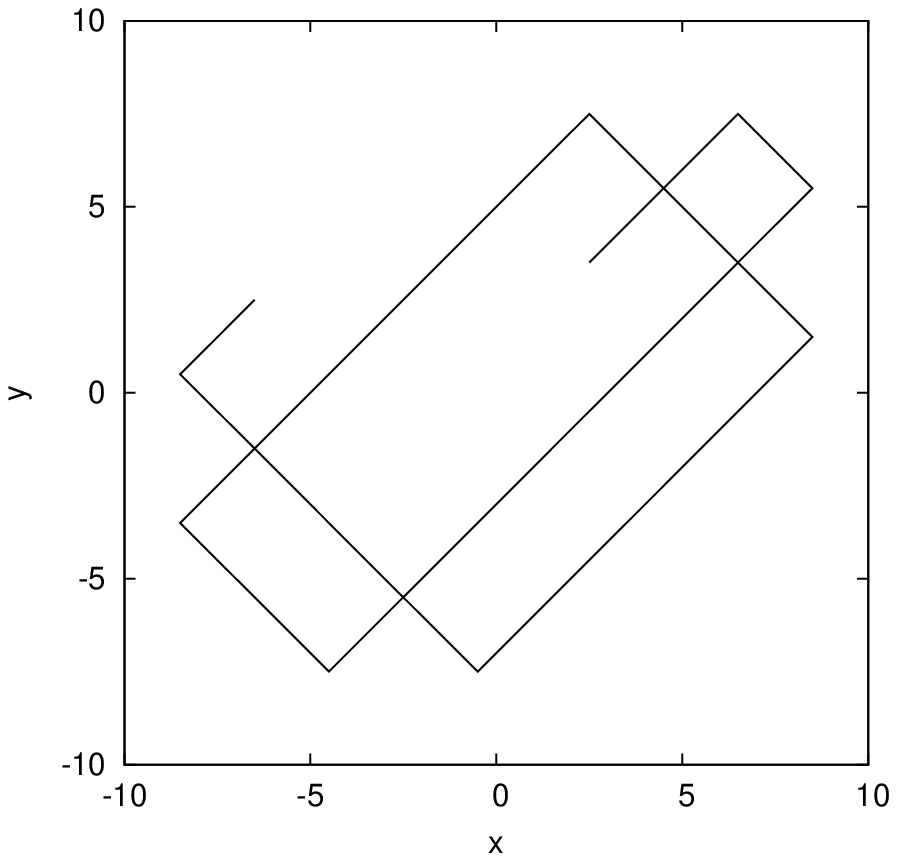}
\includegraphics[width=0.3\textwidth]{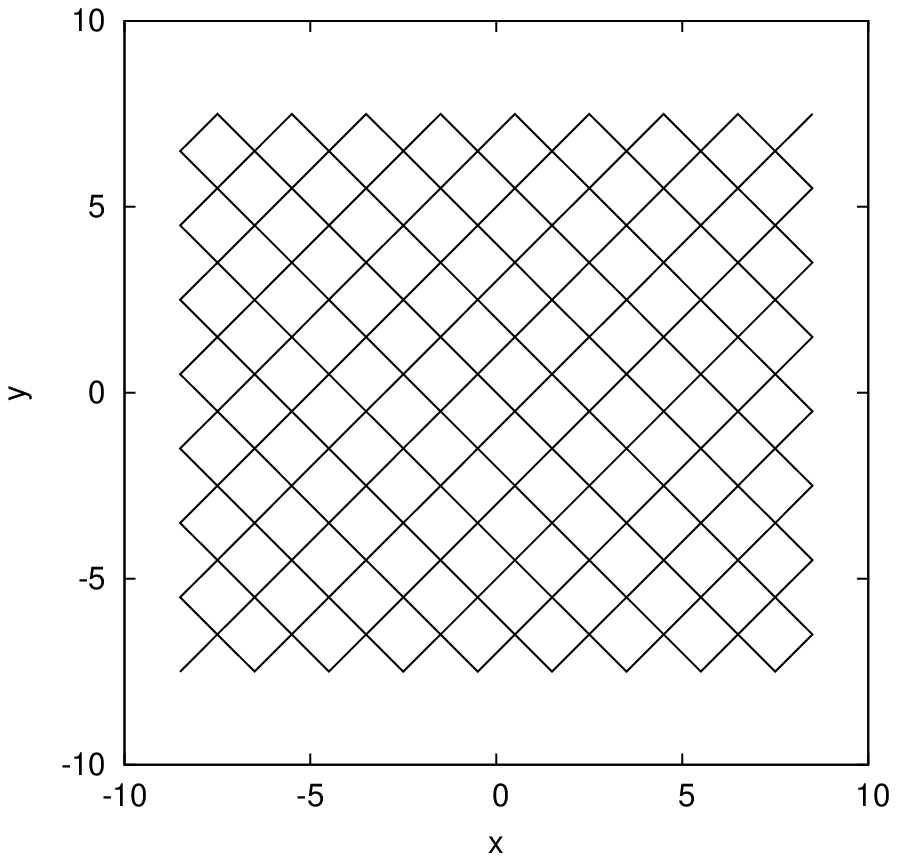}
\includegraphics[width=0.3\textwidth]{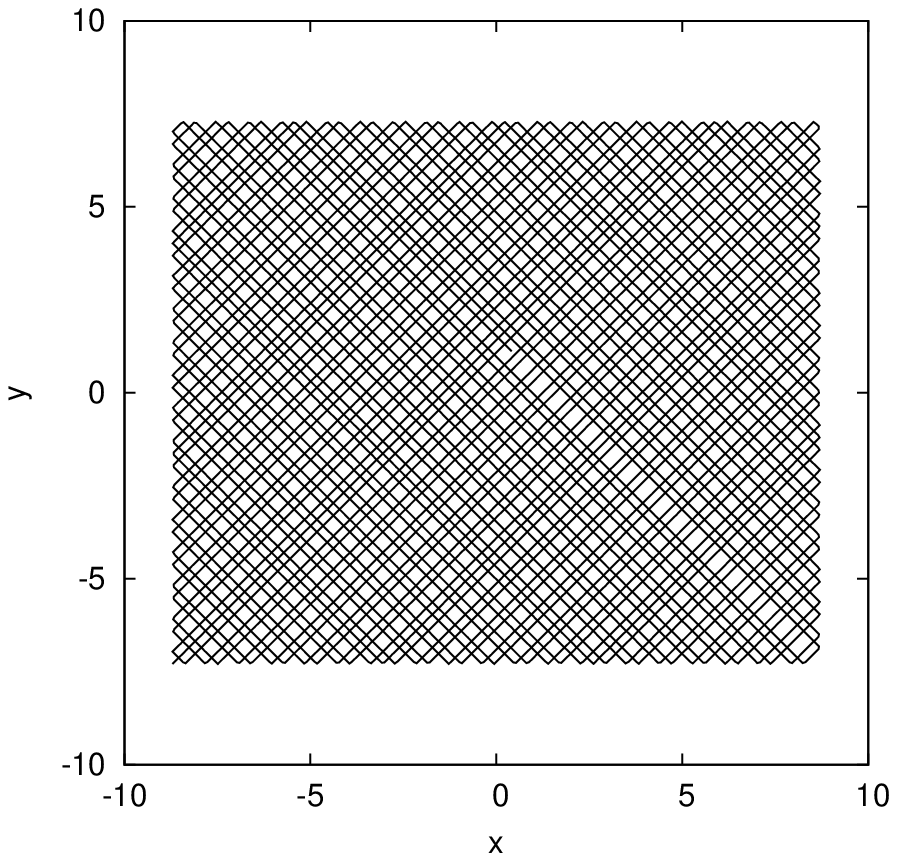}
\caption{Billiard Patterns. A small snipplet (left), a closed pattern (center) and an open pattern after some time (right). }
\label{fig:billiard}
\end{figure}

Similarly to Lissajous patterns, the bouncing billiard ball produces repeating shapes when there is a rational relation between the effective $x$ and $y$ frequencies ($f_x = 2a / v \cos \theta$ and $f_y = 2b / v \sin \theta$ in terms of the box sides $a$ and $b$ and angle $\theta$), and non-repeating, open patterns otherwise. Implementation to date use closed patterns only. Yet the non-closing patterns spread the source signals more widely, and into the higher frequencies, thus providing better isolation against $1/f$ noise, and improved noise-immunity for the larger source scales. Open billiard patterns should therefore be preferred.

Like the Lissajous patterns, the billiard scan is easily sized to preference. Its linear strokes are acceleration-free during travel across the field, but the sharp turnarounds at the edges can be mechanically demanding for telescope implementations. The main advantages of the pattern are its constant scanning speed and the uniform coverage it produces for arbitrarily scaled implementations, provided the integration is taken long enough for a large number of reflections an both sides of the bounding box.

\subsection{Spirals}

The APEX telescope\cite{apex} offers Archimedian spirals as one of two basic observing patterns available (the other being linear strokes). These spirals a defined by their constant angular and radial velocities ($\beta$ and $v_r$ respectively), and the motion is best described in polar coordinates $(r, \phi)$ where $r(t) = r_0 + v_r t$, and $\phi(t) = \phi_0 + \beta t$. The maximum radius $r_{\rm max}$ the spiral reaches is controlled by the finite time of a single pattern.

\begin{figure}[!htbp]
\centering
\includegraphics[width=0.3\textwidth]{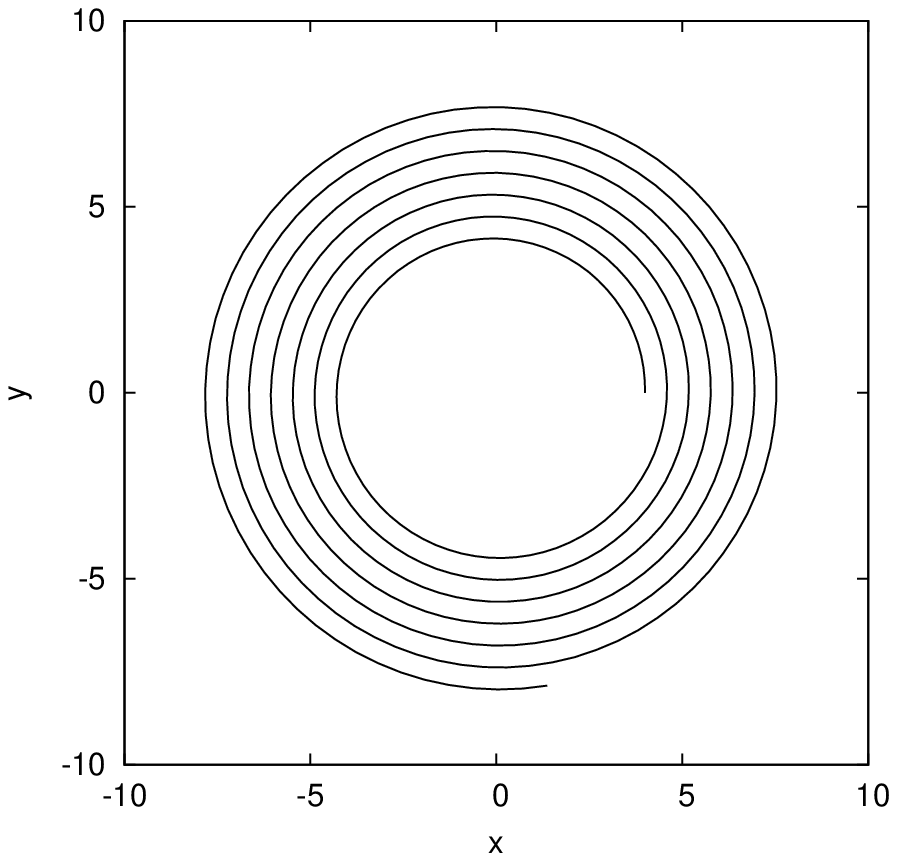}
\includegraphics[width=0.3\textwidth]{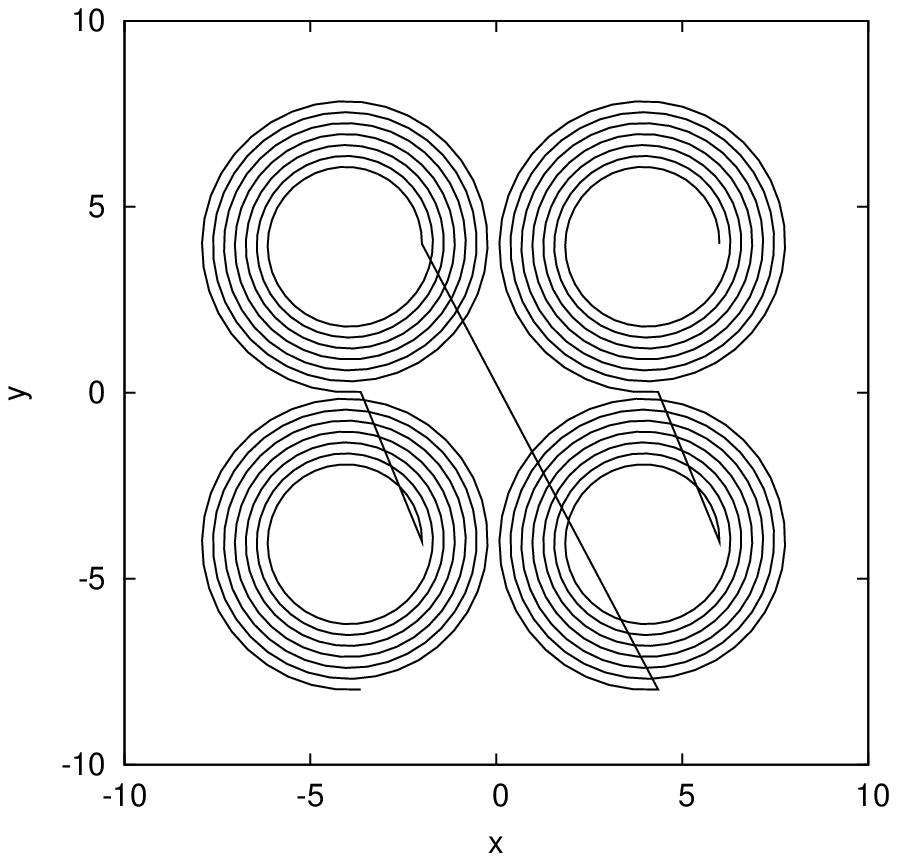}
\caption{Spirals. A single spiral (left) and a raster of smaller spirals covering the same field (right).}
\label{fig:spiral}
\end{figure}

For typical implementations, where the radial motion is slow relative to the angular one, the effective scanning velocity is approximately given by $v(t) = \beta r(t)$, and it is therefore time-variable. Thus, practical patterns grow just a few times in radius for keeping scanning speeds near their optimal value. Accordingly, in the simulations we used $r_{\rm max}/r_0$ ratios of 2, while constraining the mean scanning speed at 1 pixel/frame as previously discussed.

Spirals are good patterns that are only ever so slightly inferior to the above discussed Lissajous and billiard patterns, in giving a scant bit less noise resistance than those, especially for $1/f$ time noise (moments $m_1$ and $m_2$). Like Lissajous patterns, their trajectories are smooth and therefore well suited for mechanical telescope implementations. The pattern is also scalable to arbitrary sizes. However, the practical requirements on the spanned radial range make single spirals not suited for large field coverage, as the center parts would typically remain unexposed. 

This problem of lacking central coverage can be easily overcome by combining several spiral patterns in a raster. An example of a small raster of downsized spirals, covering the test area, is shown to the right on Figure~\ref{fig:spiral}. Such patterns are commonly used for imaging with LABOCA.

\section{LARGE FIELDS}

Some patterns, like OTF, or billiard scans, can be scaled to arbitrary sizes. Rasters of smaller patterns may also be used to cover large areas. What is the best strategy then to observe large fields -- is it better to go for the entire field at once, or mosaic it together from a raster of smaller fields? 

To answer this, we have extended the simulations to include covering with scanning patterns a large field of 64$\times$64 detector pixels. We used both enlarged OTFs and billiard scans, and also small FoV-sized patterns mosaiced together on raster of 8$\times$8 positions 16-pixels apart. From the point of view of phase-space occupation (i.e.~measured by the moments $m_0$--$m_2$), neither the up-sizing of patterns, nor the rasterization of these produced any significantly different values than what we already obtained for the smaller, or standalone, versions of the patterns. This means, that in terms of noise resistance the two strategies (going large or mosaicing) are effectively equivalent, and the observed differences are mainly caused by the choice of the underlying pattern. Thus, the results of Table~\ref{tab:summary} are applicable for large fields (both scaled and rasterized) as well as for small ones (within the estimated few percent relative systematic measurement error of the simulations).

This is not to say that other consideration cannot play a role. Larger patterns may be more easily realized with higher scanning speeds, while patterns that use secondary wobblers may be limited in range and therefore require mosaicing for large-field coverage. Thus a practical advantage may be given to one approach or the other.

\section{CONCLUSIONS}

We have used objective criteria, such as resistance to noise measured by phase-space moments, and short-timescale scanning ranges as a indicators of large-scale sensitivities, to compare some of the commonly used observing strategies for astronomical (esp.~submillimeter) imaging. 

Accordingly the best observing strategy is to move source signals randomly across the imaging array. Such modes are not easily implemented by scanning telescopes, which typically move along smooth, connected trajectories. For them, Lissajous or {\em billiard} patterns offer a reasonable compromise for mapping small and/or large fields. Spirals also exhibit formidable qualities, both as standalone patterns and when combined in rasters. However, the more traditional OTF and position-switched (chopped) modes make relatively poor choices because of their more limited abilities to stand up to adverse interference from noise, and because they can be strongly directional in their sensitivities to the larger scales, unless appropriately cross-linked at an angle or several angles. The DREAM pattern, destined for SCUBA-2, proved surprisingly weak in the simulations, both in its noise immunity and sensitivities to large scales. Finally, stare modes are most easily ruined by noise and require significant overheads in observing time when compared to the scanning modes.

Observing strategies that require secondary reflector movement are not always effective. In ground-based submillimeter applications these modes produce strong signals that result from small changes in the primary illumination. Stare modes are similarly handicapped is such applications due to the optical loading changes between dark-frame calibrations and on-source observing, not to mention the significantly longer observing times that stare modes can require to reach the comparable sensitivities to scans.

To retain optimal sensitivities on the large scales, scanning patterns need to be comparable in size to the measured scales. Large fields may be mapped either by scaled versions of the patterns (where applicable), or by mosaicing together smaller fields. The two strategies are similar in their characteristic resistances to noise, which is mainly defined by the type rather than the size of the pattern used.

\acknowledgements

The author wishes to thank Colin Borys and Darren Dowell, for their help in designing better observing modes for SHARC-2, and Hiroshige Yoshida for promptly implementing these at the CSO. The author also recalls many spiralling discussions over coffee with Axel Weiss on suitable patterns for imaging with the bolometer arrays at APEX.

\bibliographystyle{spiebib}   

\begin{table}[tbp]
\centering
\begin{tabular}{llcccll}
\hline
              & \multicolumn{1}{c}{\bf Geometric} &  \multicolumn{3}{c}{\bf Moments} & &          \\
\cline{3-5}
{\bf Pattern} & \multicolumn{1}{c}{\bf Paramters} & $m_0$ & $m_1$ & $m_2$ & \multicolumn{1}{c}{$l_c$} & {\bf Comments} \\
\hline
\hline
random             & $a,b$      &
                                   1.000     & 1.000    & 1.000 & $a, b$ & discrete, unfeasible(?) \\
Lissajous          & $A_x, A_y, \omega_y/\omega_x$   &
                            0.129  & 0.126 & 0.125  & $2A_x, 2A_y$ & smooth \\
billiard (open)    & $a, b, \theta$   &
                            0.097  & 0.089  & 0.086 & $a, b$ \\
billiard (closed)  & (see above)  &
                            0.091  & 0.068  & 0.058 & $a, b$ \\
rotating OTF       & $L, \Delta, \delta \Theta$ &
                            0.088  & 0.085  & 0.084  & $L$ & requires several angles 0--90$^\circ$ \\
raster of spirals  & $\Delta_{\rm ras}, r_0, r_{\rm max}$ &
                            0.080  & 0.073  & 0.070 & $2r_{\rm max}$ \\
spiral             & $r_0, r_{\rm max}$ &
                            0.061  & 0.056  & 0.054  & $2r_{\rm max}$ & smooth \\ 
crossed OTF (90$^\circ$)        & $L, \Delta$ &
                            0.035  & 0.035  & 0.035  & $L$ \\
chop               & $d$ &
                            0.030  & 0.030  & 0.045  & $d$ & discrete, (oriented), secondary \\
OTF                & $L, \Delta$ &
                            0.018  & 0.018  & 0.018  & $\Delta, L$ & strongly oriented \\
DREAM              & &
                            0.018  & 0.018  & 0.019  & 4 pixels & discrete, secondary \\
stare              & &
                            n/a    & 0.000  & 0.000  & FoV & up to 4$\times$ integration time \\      
\hline
\\
\end{tabular}
\caption{Comparison of popular scanning modes ranked by their generic noise resistance (as measured by $m_0$). The higher moments ($m_1$ and $m_2$) represent the relative resistance to $1/f$ and $1/f^2$ type noise respectively, and also hint at the noise resistance of the more extended source scales. The column $l_c$ is the largest-scale structure that can be recovered under the white-noise sensitivity limit (see~Eq.~\ref{eq:extended-noise}). When the critical scale $l_c$ is direction-dependent, a comma-separated range of typical values are indicated. Random patterns are best in theory, however, these cannot be implemented with telescopes that move along smooth trajectories. Random patterns, however, maybe possible for scanning in frequency space (e.g.~for spectroscopic receivers). 
}
\label{tab:summary}
\end{table}

\begin{figure}[p]
\includegraphics[width=0.3\textwidth]{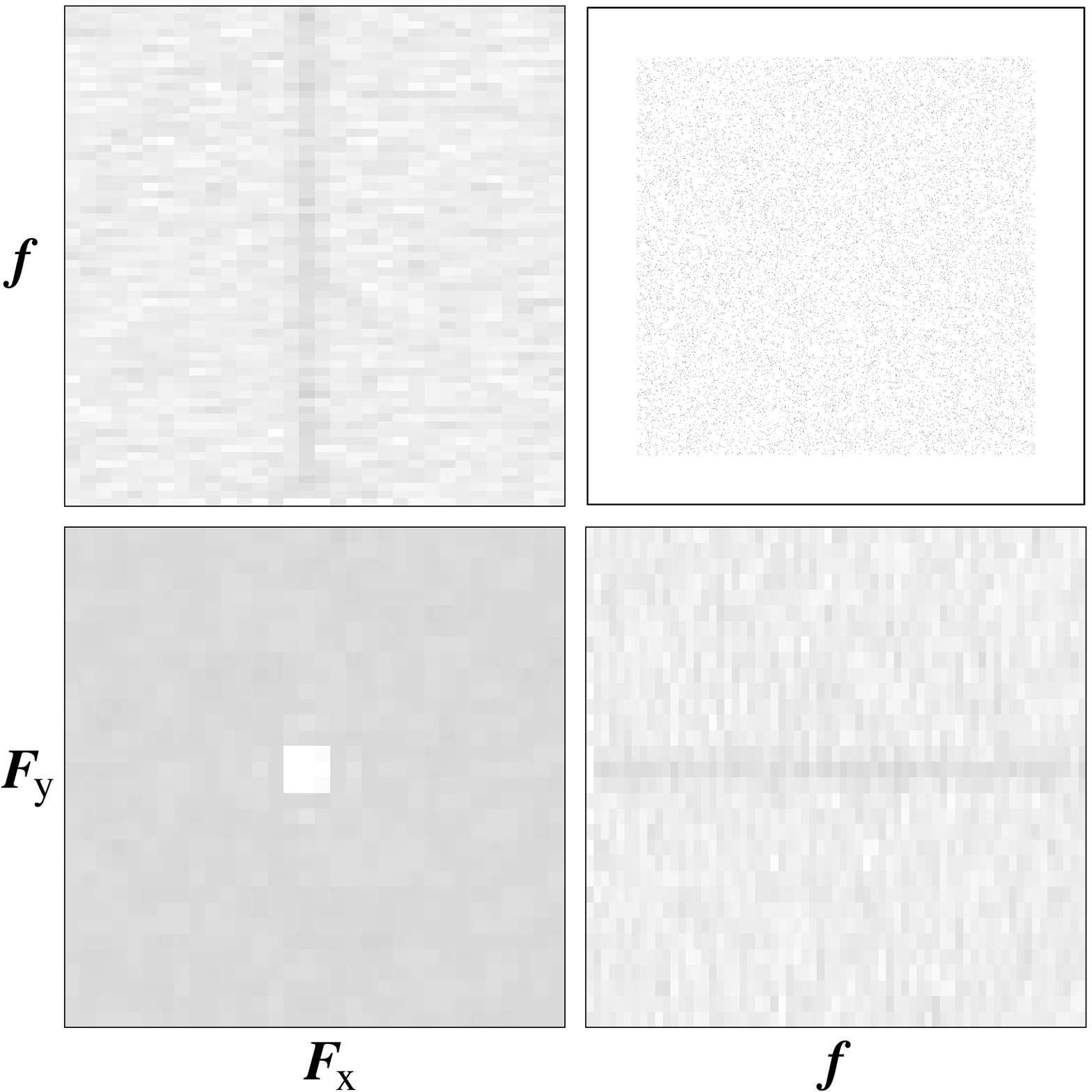}
\hfill
\includegraphics[width=0.3\textwidth]{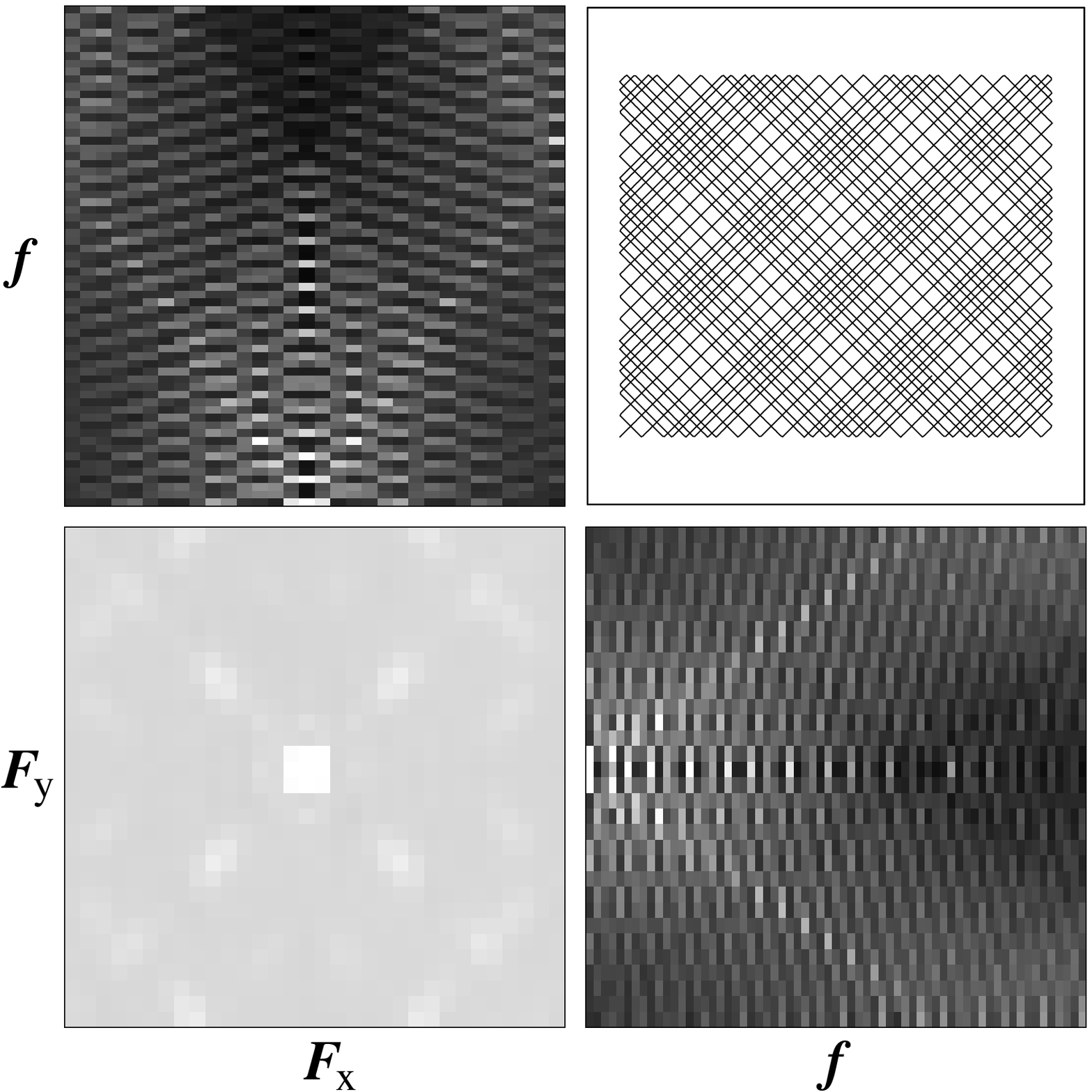}
\hfill
\includegraphics[width=0.3\textwidth]{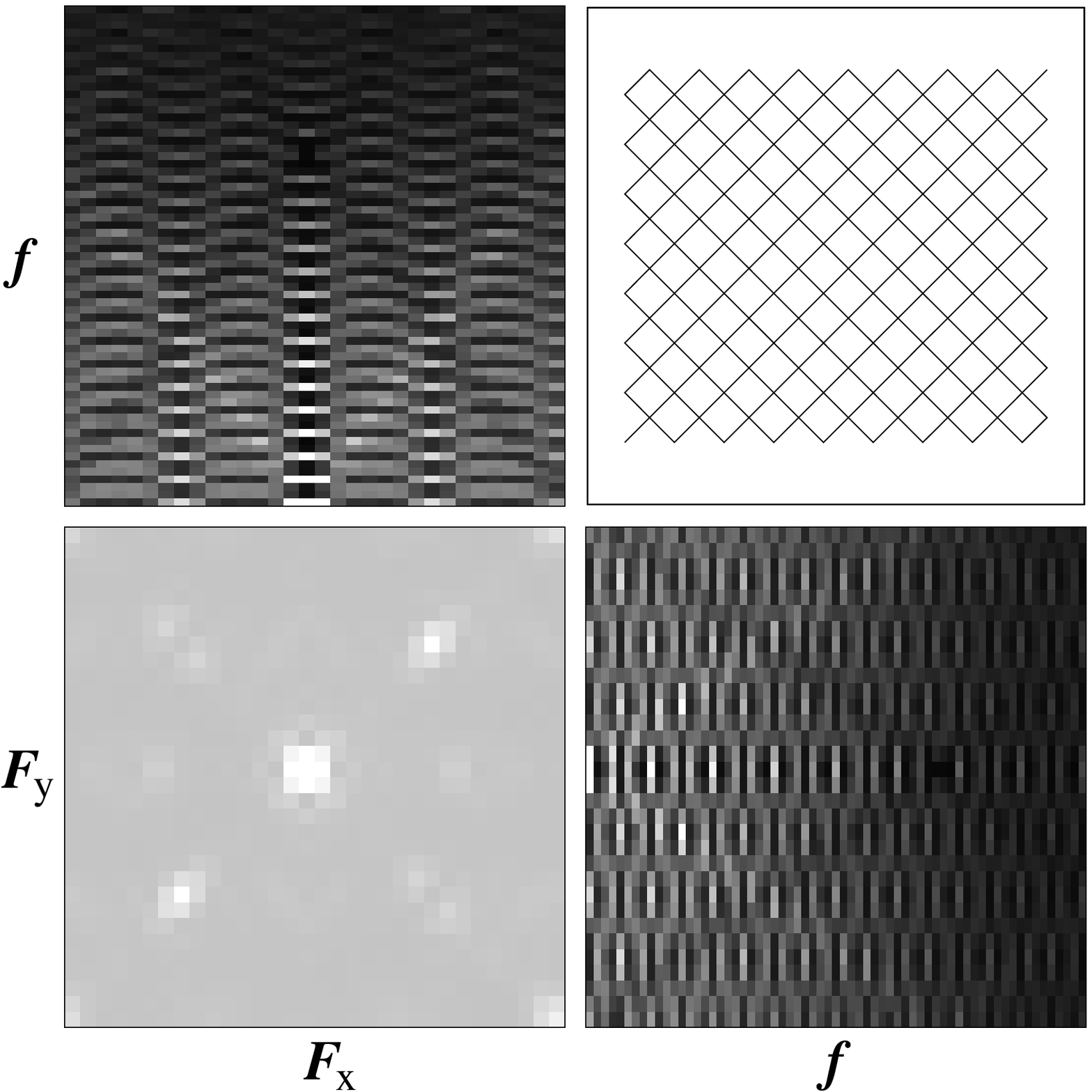}
\vskip 12pt
\includegraphics[width=0.3\textwidth]{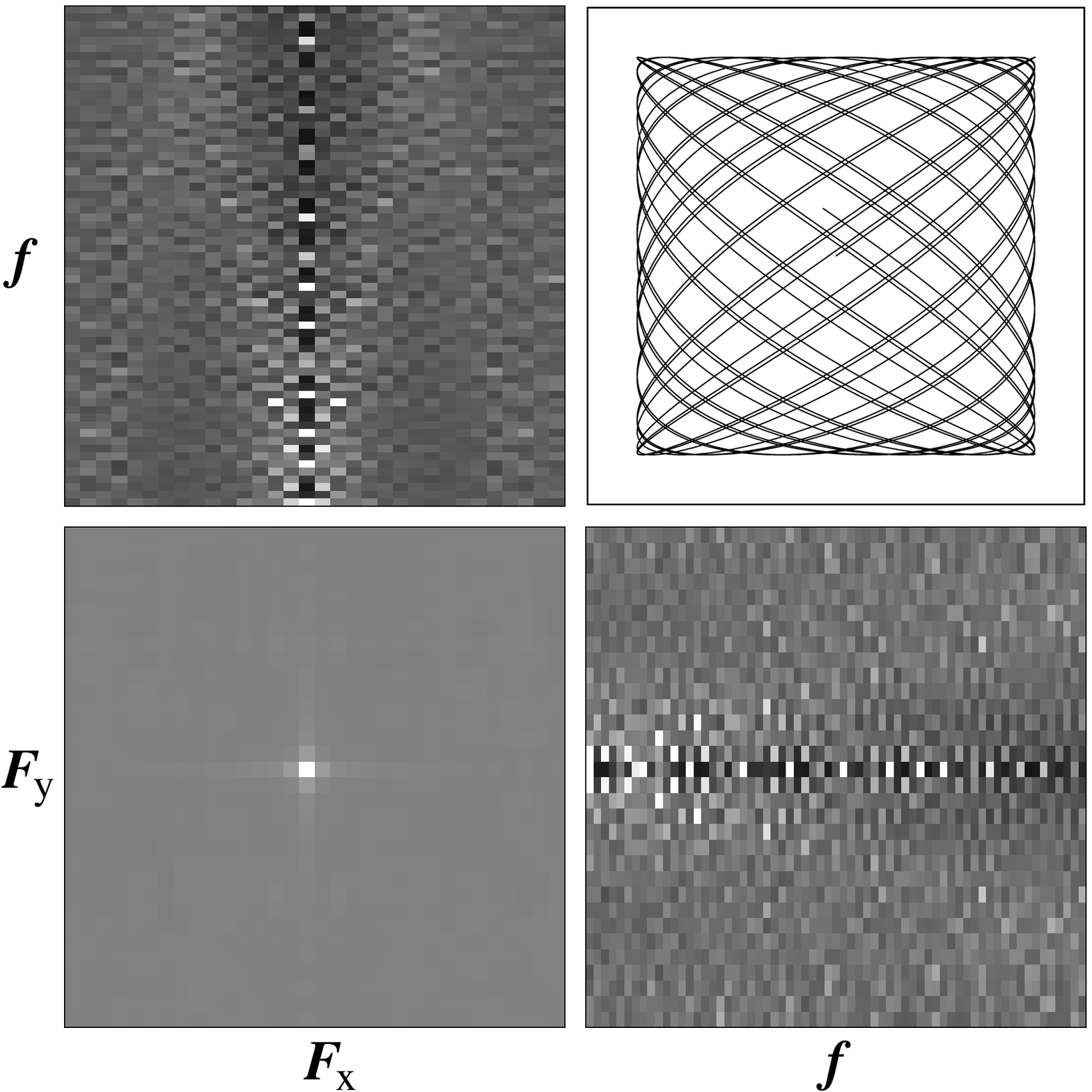}
\hfill
\includegraphics[width=0.3\textwidth]{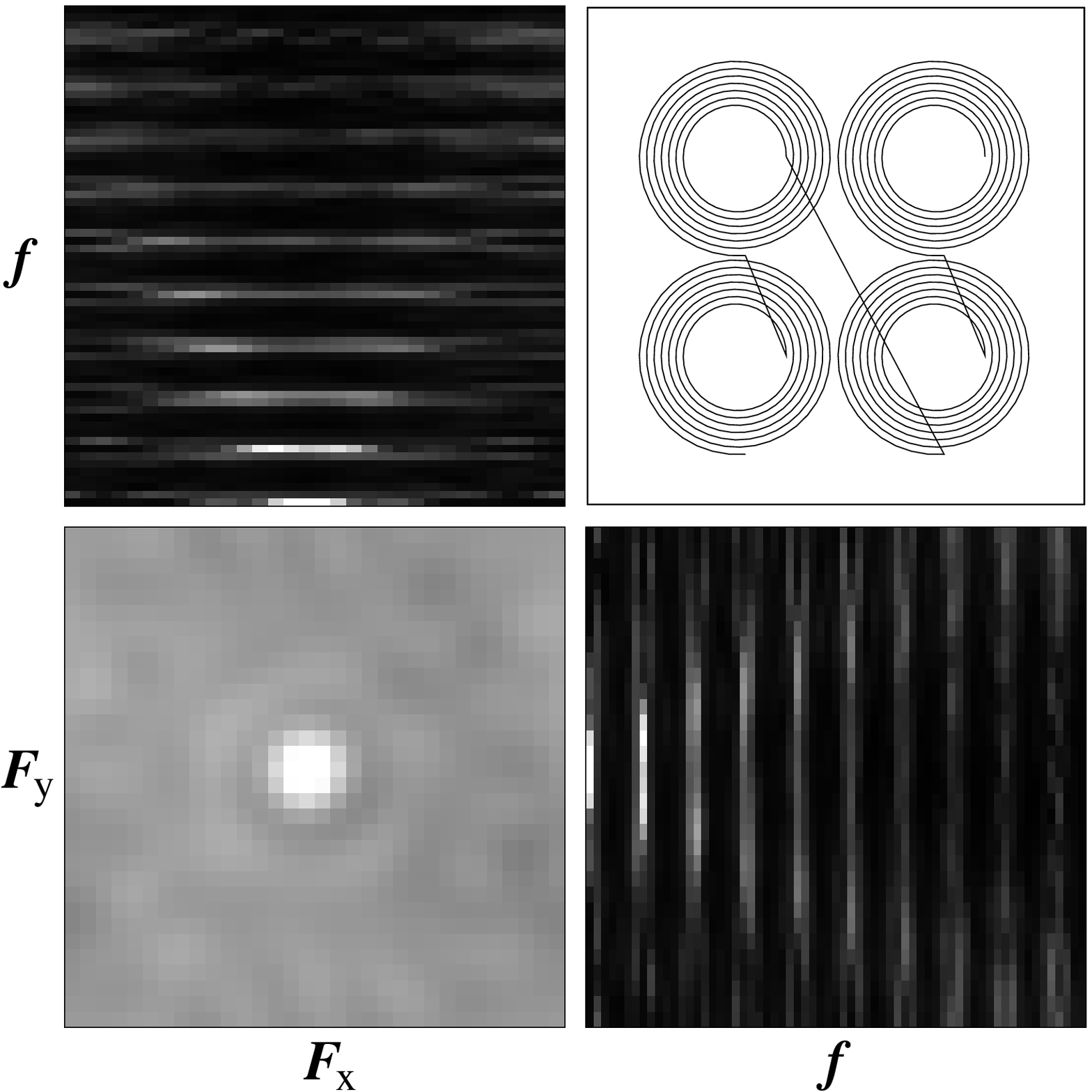}
\hfill
\includegraphics[width=0.3\textwidth]{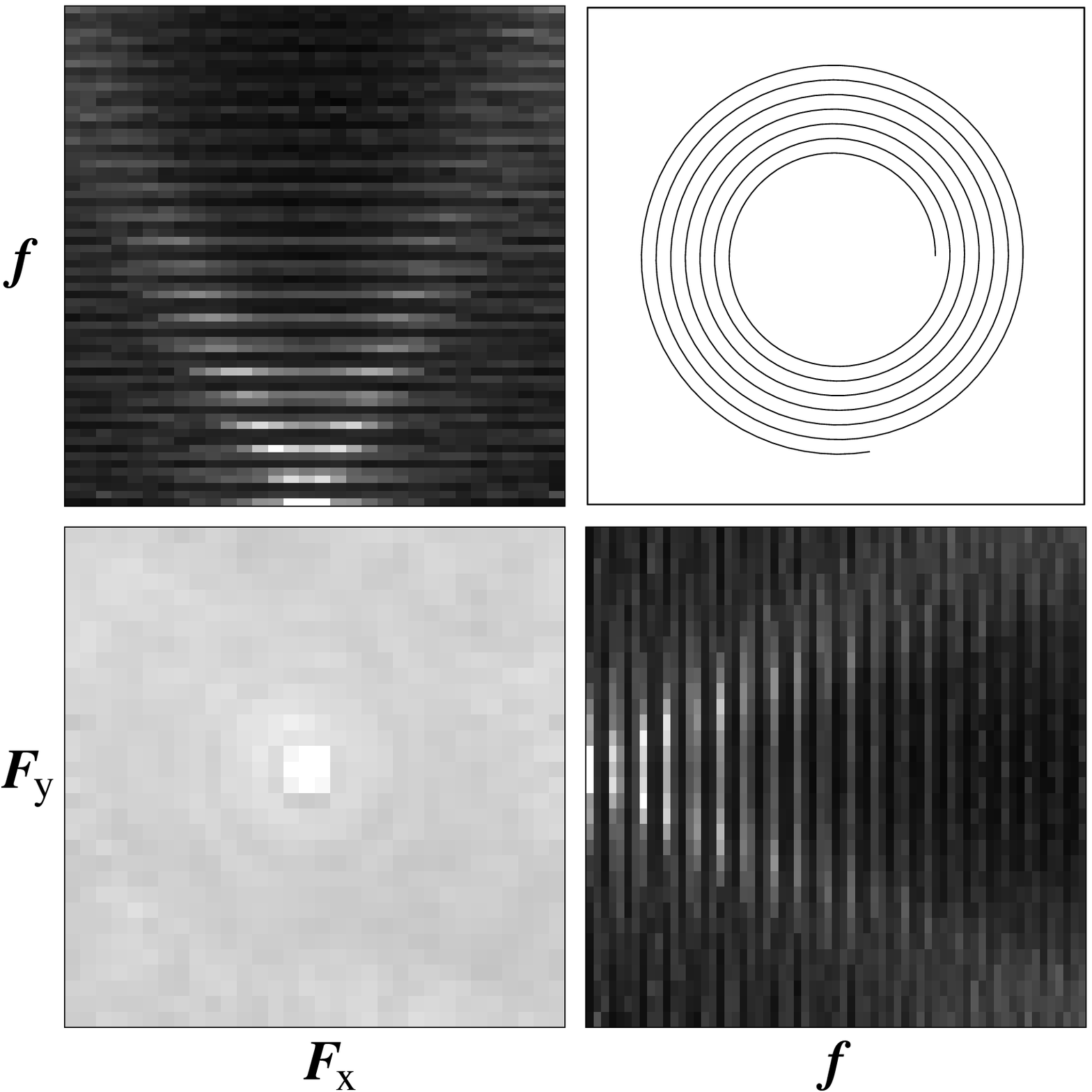}
\vskip 12pt
\includegraphics[width=0.3\textwidth]{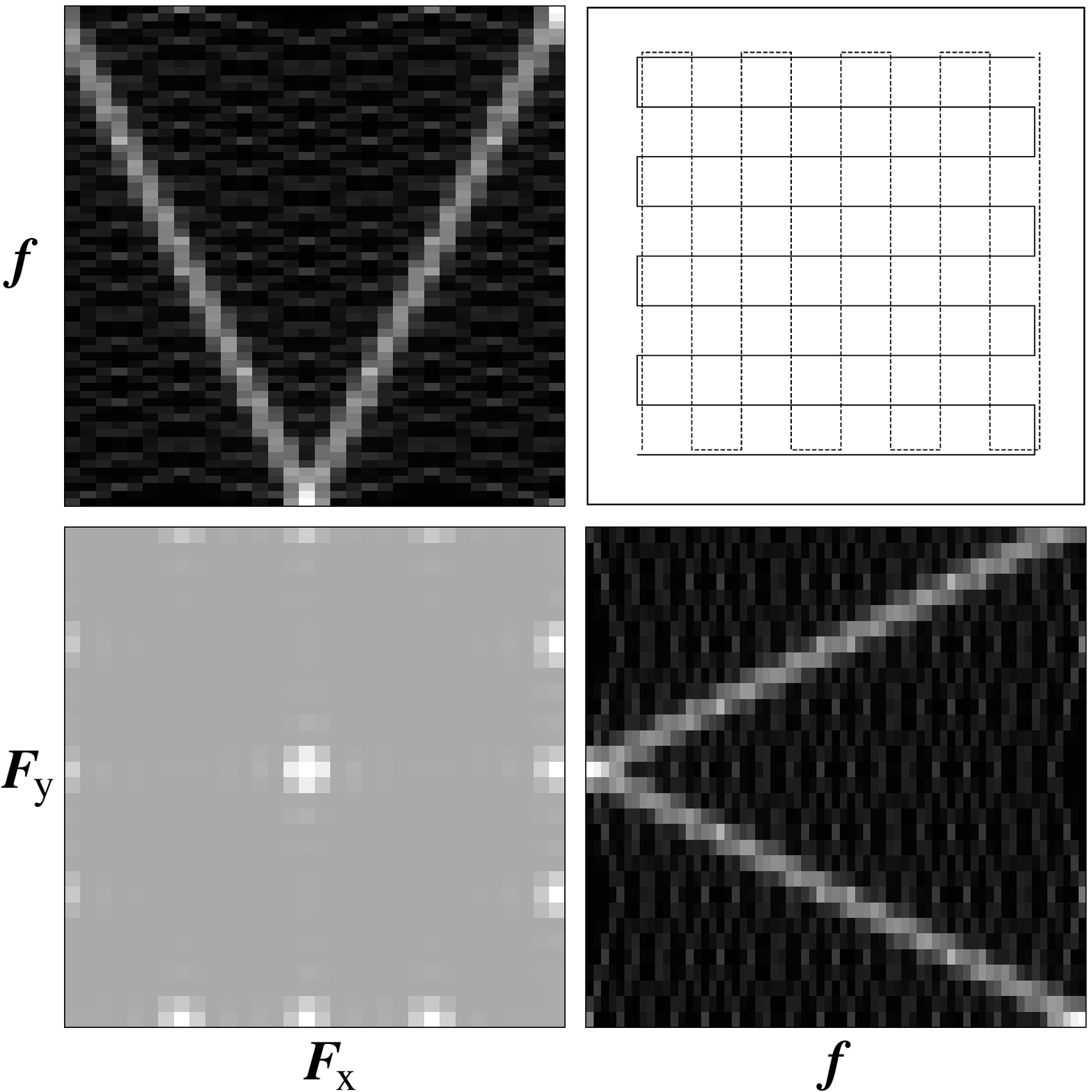}
\hfill
\includegraphics[width=0.3\textwidth]{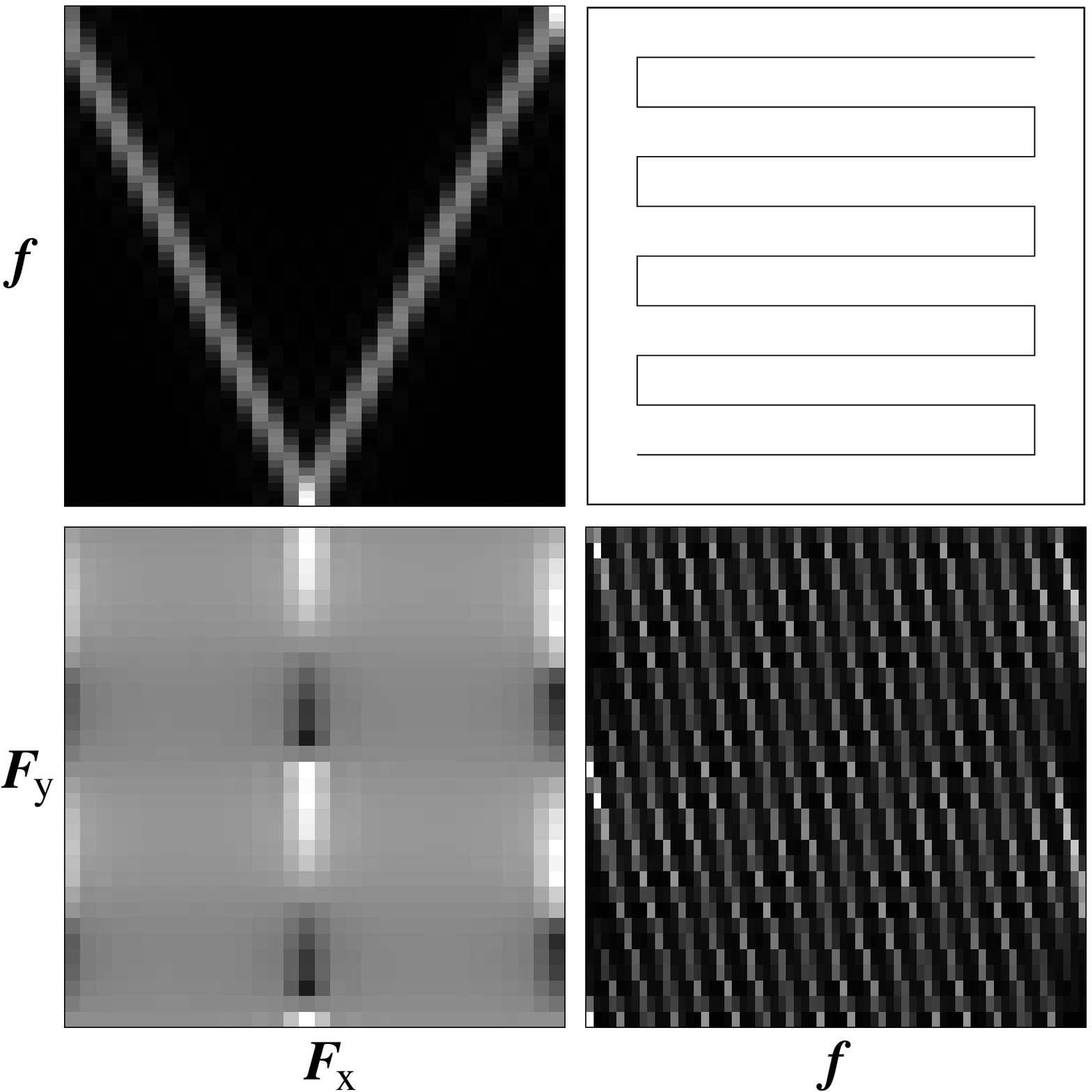}
\hfill
\includegraphics[width=0.3\textwidth]{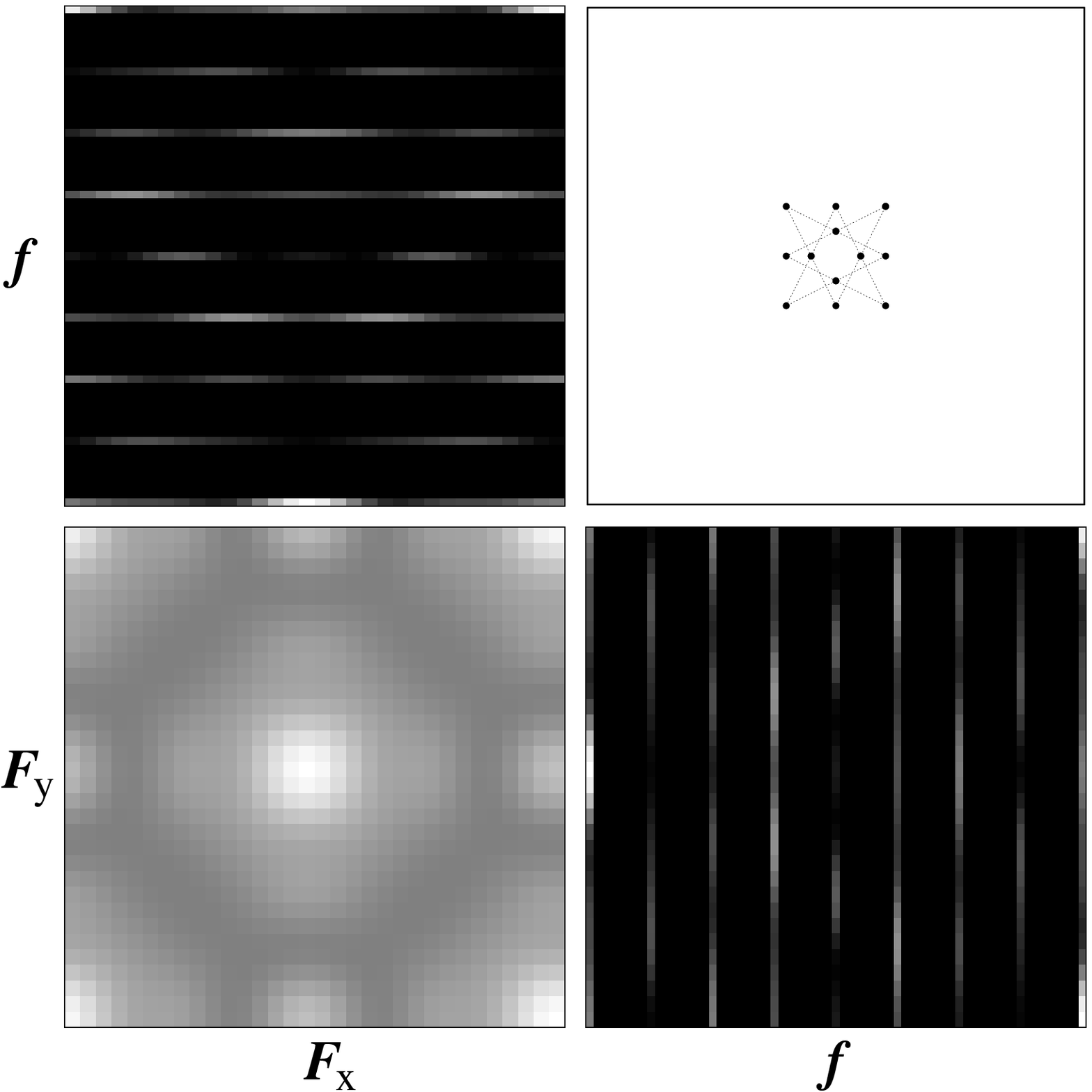}
\vskip 12pt
\caption{
Point-source spectral projections for a few observing modes. Top row (left to right): random, open-, and closed billiard patterns. Middle row: Lissajous, raster of spirals, single spiral. Bottom row: cross-linked OTF, single-pass OTF, and DREAM. The greyscale images are autoscaled between zero (black) and their maximum values (white) outside of the zero frequency bins.
}
\label{fig:spectra}
\end{figure}

\end{document}